\documentclass[lettersize,journal]{IEEEtran}
\usepackage{amsmath,amsfonts,amsthm}
\usepackage{algorithmic}
\usepackage{array}
\usepackage[caption=false,font=normalsize,labelfont=sf,textfont=sf]{subfig}
\usepackage{textcomp}
\usepackage{stfloats}
\usepackage{url}
\usepackage{verbatim}
\usepackage{graphicx}
\usepackage{cite}
\usepackage{xcolor}
\usepackage{orcidlink}
\usepackage{hyperref}
\usepackage{cuted, lipsum}
\usepackage{accents}
\usepackage[nocomma]{optidef}
\usepackage{stfloats}
\usepackage{lipsum}
\usepackage{mathrsfs}
\usepackage{nicefrac}
\usepackage{xfrac}

\def\BibTeX{{\rm B\kern-.05em{\sc i\kern-.025em b}\kern-.08em
    T\kern-.1667em\lower.7ex\hbox{E}\kern-.125emX}}
\usepackage{balance}
\newcommand{\contradiction}{%
  \ensuremath{{\Rightarrow\mspace{-2mu}\Leftarrow}}%
}

\begin{document}

\title{A Control Architecture for Fast Frequency Regulation with Increasing Penetration of Inverter Based Resources}
\author{Jose A. Solano-Castellanos \orcidlink{0009-0006-8580-4180}, \textit{Student Member, IEEE}; Hassan Haes Alhelou \orcidlink{0000-0001-8176-1589}, \textit{Senior Member, IEEE}; Ali T. Al-Awami, \textit{Senior Member, IEEE} \orcidlink{0000-0003-0062-2013}; 
Mohannad Alkhraijah \orcidlink{0000-0003-2116-6766} \textit{Member, IEEE}; and Anuradha M. Annaswamy \orcidlink{0000-0002-4354-0459}, \textit{Life Fellow, IEEE}  
\thanks{This work was supported by the Center for Complex Engineering Systems (CCES) at the King Abdulaziz City for Science and Technology (KACST) and the Massachusetts Institute of Technology (MIT) (Corresponding author:
Jose A. Solano-Castellanos).}
\thanks{Jose A. Solano-Castellanos and Anuradha M. Annaswamy are with the Department of Mechanical Engineering, Massachusetts Institute of Technology, Cambridge, MA 02139, USA. (email: jsolanoc@mit.edu; aanna@mit.edu).}
\thanks{Hassan Haes Alhelou is with the School of Engineering, Massachusetts Institute of Technology, Cambridge, MA 02139, USA (e-mail: alhelou@mit.edu).}
\thanks{Ali T. Al-Awami is with the Department of Electrical Engineering and the
Interdisciplinary Research Center for Smart Mobility and Logistics, King Fahd University of Petroleum \& Minerals, Dhahran 31261, Saudi Arabia (e-mail: aliawami@kfupm.edu.sa).}
\thanks{
Mohannad Alkhraijah is with the Smart Cities Institute, King Abdulaziz City for Science and Technology, Riyadh 12354, Saudi Arabia (e-mail: malkhraijah@kacst.gov.sa).}
}


\maketitle


\begin{abstract}
This paper addresses frequency regulation under operational constraints in interconnected power systems with high penetration of inverter-based renewable generation. A two-layer control architecture is proposed that combines optimized droop and Virtual Synchronous Machine (VSM) primary control with a Model Predictive Control (MPC) secondary layer operating at realistic control-room update rates. Unlike recently proposed approaches, the proposed framework integrates MPC within existing grid control structures, enabling constraint-aware coordination. A reduced-order frequency response model is systematically derived from a detailed grid model using Hankel singular values, and a reduced-order Kalman–Bucy observer enables state and disturbance estimation using only measurable outputs. Evaluation using representative data from the Kingdom of Saudi Arabia demonstrates effective frequency regulation under realistic operating conditions.
\end{abstract}

\begin{IEEEkeywords}
Hankel Singular Value; Inverter-Based Resources; Low-Inertia Systems; Model Predictive Control; Multi-Area Power Grid; Optimal Control; System Frequency Response.
\end{IEEEkeywords}


\section{Introduction}

The growing concern over greenhouse gas emissions and the lack of sustainability of fossil- and carbon-based power plants have led nations around the world to opt for renewable energy sources (RESs). Ambitious projects around the world that align with a greener future have led to a large-scale penetration of inverter-based resources (IBRs). However, a high share of IBRs imposes new challenges on the operation, control, and resilience of power systems. Unlike traditional Synchronous Generators (SGs), IBRs lack rotational inertia that leads to larger rates of change of frequency (RoCoF) and frequency deviations 
\cite{Kavvathas2026NovelCharacteristics}, which can trigger protective mechanisms such as load-shedding or even cause large-scale blackouts 
\cite{Alhelou2024DynamicSystems}. To mitigate these undesirable outcomes, methodologies that achieve fast frequency regulation and synthetic inertia are needed, which are often denoted as 
Fast Frequency Control Ancillary Services \cite{Alhelou2024AustraliasDirections}, or Fast Frequency Response \cite{Meng2020FastIssues}.

As time constants shorten with the increased penetration of IBRs and uncertainty in power generation grows due to the variability of natural resources, there is a pressing need to control and coordinate IBRs and SGs using fast-acting control mechanisms that explicitly account for energy storage limitations, grid code requirements, and the coexistence of dispatchable and non-dispatchable RESs. This paper therefore pertains to a control architecture that provides fast frequency regulation in a power system with a large penetration of RESs grid-connected through IBRs.

Although IBRs lack the inherent stabilizing effect of mechanical inertia, they provide the requisite compensation through fast and flexible response. Grid-forming voltage source converters (VSCs) with DC-side energy buffers can adjust power output to frequency deviations far faster than SGs \cite{Stanojev2022MPC-BasedSystems, Rocabert2012ControlMicrogrids}. SGs’ mechanical power compensation has time constants of $3-10$ seconds (hydro to steam turbines), whereas VSCs respond in $5-500$ milliseconds, depending on the source \cite{Dorfler2023ControlSystems}.
The control architecture that we propose in this paper leverages the fast regulation capability of IBRs and control approaches that combine droop \cite{Chandorkar1993ControlSystems}, Virtual Synchronous Machine (VSM) \cite{Beck2007VirtualMachine, Zhong2011Synchronverters:Generators}, and Model Predictive Control (MPC) \cite{Richalet1978ModelProcesses, Mayne2000ConstrainedOptimality}.

Droop-based and VSM are the two most widespread 
control approaches in the literature and implemented in the industry, both of which aim to emulate characteristics of SGs. Droop-based control uses the traditional droop approach in SGs to regulate active and reactive power at the inverter's output. VSMs on the other hand aims to emulate the swing equation of SGs and effectively increase the inertia in the grid. However, the storage capacity of VSCs, in the form of the charge stored in the DC capacitor, is much smaller than that of SGs which is in the form of kinetic energy stored in the rotor \cite{Dorfler2023ControlSystems}.
As pointed out in \cite{Ashabani2014NovelSynchronous-VSC}, this results in significant constraints on the achievable VSM and droop-based gains. These limitations, together with increasingly stringent grid-code requirements on frequency deviations, RoCoF, and power exchange limits, motivate the adoption of control strategies that go beyond linear controllers. In this context, MPC has emerged as a promising framework for frequency regulation in modern power grids. MPC is an optimization-based, discrete-time control strategy that explicitly incorporates system dynamics, operational constraints, and performance objectives into a receding-horizon optimization problem. Over the last decade, MPC has gained significant attention in power system applications due to its ability to handle constraints on input magnitudes, input rates, and system states, while providing fast and flexible control actions that can adapt to changing operating conditions.

In the context of control of power systems, MPC has been explored in \cite{Stanojev2022MPC-BasedSystems, Ersdal2016ModelControl,Hu2024ResilientAttacks, Liu2024Tube-BasedPenetration, Zhang2023RobustError, Yang2021DistributedStorage, Ademola-Idowu2021FrequencySystems}. In \cite{Ersdal2016ModelControl}, 
MPC was formulated for Automatic Generation Control (AGC) where it was shown to improve frequency regulation and robustness against uncertainties when compared to classical PID control. The approach in \cite{Ersdal2016ModelControl} consisted of a centralized non-linear MPC framework for AGC in the Nordic power system that took into account constraints on tie-line power flow, generation capacity, and generation rate of change. However, this approach does not account for the consequences of integrating IBRs into the grid. 

The approaches in \cite{Stanojev2022MPC-BasedSystems} and \cite{Ademola-Idowu2021FrequencySystems} employed MPC to deliver fast frequency control in converter-dominated power grids. In \cite{Hu2024ResilientAttacks}, the authors propose MPC-based approaches for load frequency regulation under zero-mean Gaussian disturbances that leverage wind turbines for frequency support. The authors of \cite{Stanojev2022MPC-BasedSystems} incorporated converter constraints such as limits on allowable frequency excursions to prevent tripping, as well as battery state-of-charge restrictions, directly within the MPC formulation. A parallel line of work has explored distributed and robust formulations to handle model uncertainty directly within the MPC constraints: \cite{Liu2024Tube-BasedPenetration} employs a tube-MPC scheme combining a nominal controller with an online correction term to bound the effect of wind-power disturbances and linearization error, while \cite{Zhang2023RobustError} formulates an MPC-adjacent numerical-optimal-control formulation that enforces constraints under a worst-case disturbance scenario. Similarly, distributed MPC formulations coordinate multiple areas or units through consensus or projection-based algorithms \cite{Yang2021DistributedStorage}.

Despite the promising results reported in \cite{Stanojev2022MPC-BasedSystems, Ersdal2016ModelControl,Hu2024ResilientAttacks, Liu2024Tube-BasedPenetration, Zhang2023RobustError, Yang2021DistributedStorage, Ademola-Idowu2021FrequencySystems}, some challenges associated with low-inertia grids remain unsolved. The required control update rates, ranging from 50 milliseconds to 1 second, are too fast for practical implementation, as they are well below the typical 2-6 seconds update rate of practical SCADA systems \cite{Karvelis2022StateLinks}. Distributed approaches like \cite{Yang2021DistributedStorage} require bidirectional peer-to-peer communication between neighboring units and, as the authors note, incur a convergence delay for the consensus algorithm as well as potentially heavy aggregate computational burden relative to conventional AGC. This requirement is difficult to reconcile with current grid operational infrastructure, where individual generation and converter units communicate only with central SCADA systems rather than directly with one another; a distributed architecture of this kind would therefore require substantial changes to existing communication infrastructure to be deployed in practice. 

In terms of constraint handling, the robust formulation of \cite{Zhang2023RobustError} relies on reachability tools, resulting in solution times approaching 10 seconds per instance, while \cite{Liu2024Tube-BasedPenetration} relies on tube-based constraint tightening to guarantee robustness against bounded disturbances. Both approaches require a well-characterized uncertainty set or disturbance bound to construct their respective guarantees, which may not be readily available in practice, particularly for renewable-driven disturbances whose statistics can vary with weather conditions, time of day, and grid topology. In addition, much of the above papers rely on highly simplified turbine models or models that are only compatible with steam power plants. Even though more systematic modeling approaches have been developed, such as \cite{Zhang2024IntegratingSystems}, where a Hankel-based balanced truncation reduced-order model is obtained and later refined using a radial basis function neural network trained on measurement data, the initial model is still highly simplified and limited to a steam turbine representation.

MPC-based approaches face an additional challenge: since full-state access is not available in power systems, an observer is required to estimate the system state as well as the unknown disturbance driving the frequency deviation, and the detectability of the resulting disturbance-augmented system is seldom formally established. The control approach should be validated using a detailed model of the underlying dynamics. 

Our proposed control architecture addresses these challenges directly: it operates at a realistic SCADA-compatible control update rate requiring only the existing communication link between the control room and each region, enforces performance constraints through a slack-variable relaxation that does not require an a priori characterization of the disturbance, accommodates heterogeneous turbine types, and is evaluated using a detailed linear model. In this sense, our approach overcomes the shortcomings of \cite{Stanojev2022MPC-BasedSystems, Hu2024ResilientAttacks,Liu2024Tube-BasedPenetration,Zhang2023RobustError, Yang2021DistributedStorage, Ademola-Idowu2021FrequencySystems}. Motivated by the Vision 2030 initiative in the Kingdom of Saudi Arabia (KSA) \cite{SaudiInitiative}, we use the KSA power grid with its projected renewable energy penetration as a case study.

\subsection{Paper Contributions and Organization}

In this paper, we propose a two-layer control architecture that follows the traditional primary-secondary control hierarchy, facilitating adoption by transmission system operators while respecting existing communication and real-world industry constraints. The primary control layer is designed by optimizing the parameters of droop- and VSM-based control schemes, capturing the tradeoff between control effort, frequency deviation, and RoCoF. The secondary control layer employs an MPC scheme that explicitly accounts for frequency deviation, RoCoF, operational constraints, and the day-ahead commitment. Since the system state required by the MPC is not directly accessible, this paper proposes a reduced-order observer that systematically reduces the model dimension and enables online state and disturbance estimation.

The main contributions of the paper are as follows:
\begin{enumerate}
    \item an integrated control architecture for frequency regulation that combines MPC with existing droop- and VSM-based control schemes, specifically tailored for power grids with high penetration of IBR generation; the MPC operates at control update rates compatible with existing SCADA infrastructure and handles performance constraints through a slack-variable formulation;
    \item a modular reduced-order modeling procedure for interconnected power grids, in which Hankel singular value (HSV) truncation is performed independently on each operational region and the resulting reduced-order subsystems are assembled into a single interconnected reduced-order model of the grid;
    \item the integration of a reduced-order Kalman-Bucy observer for joint state and disturbance estimation, enabling the controller to operate using only measurable outputs, for which we establish sufficient conditions guaranteeing detectability of the disturbance-augmented system; and
    \item evaluation of the proposed approach using representative data from the KSA power grid on the derived detailed linear model, demonstrating its effectiveness under realistic operating conditions.
\end{enumerate}
The remainder of the paper is organized as follows. Section \ref{sec:Problem Statement} formulates the problem. Section \ref{sec:SFR Model} derives the system frequency response (SFR) model for an interconnected grid. Section \ref{sec:Optimal Control} presents the proposed two-layer control architecture combining droop, VSM, and MPC strategies. Section \ref{sec:Primary Control} details the optimized primary control design, while Section \ref{sec:MPC} introduces the secondary MPC-based controller for performance and constraint management. Section \ref{sec:Observer} describes the reduced-order model and observer design. Simulation results are provided in Section \ref{sec:Numerical Results}, followed by conclusions and future work in Section \ref{sec:Conclusions}.


\section{Problem Statement}\label{sec:Problem Statement}

We consider the problem of frequency regulation in an interconnected power grid with a high penetration of inverter-based renewable generation. The grid consists of $N$ operational regions indexed by $i \in \{1, \ldots, N\}$, which are interconnected through $L$ AC tie-lines indexed by $l \in \{1, \ldots, L\}$. The \textit{i}-th operational region exchanges power with a subset of neighboring regions denoted by $\mathcal{N}_i$. We assume that one or more of these regions may experience a sudden loss of generation or an increase in load, modeled as a step disturbance. The total grid disturbance is denoted by $\Delta P(t) \in \mathbb{R}^N$, whose \textit{i}-th component $\Delta P_i(t)$ represents the step disturbance in the \textit{i}-th region and is assumed to be piecewise constant. We also address the effect of a noisy pseudorandom disturbance in Section \ref{sec:Numerical Results} to expand the scope of our approach to time-varying disturbances.

This paper will focus on the following two problems:

\noindent \textbf{Problem 1:} Derive a detailed linear continuous-time state-space model of the interconnected power grid of the form
\begin{equation}\label{eq:state eq}
    \dot{x}(t) = A x(t) + B\left( \Delta P - \Lambda u(t) \right) 
\end{equation}
\begin{equation}\label{eq:meas eq}
        y(t) = C x(t)
\end{equation}
where $x(t) \in \mathbb{R}^n$ denotes the state vector, $u(t) \in \mathbb{R}^N$ represents the control input, and $y(t) \in \mathbb{R}^{N+L}$ denotes the vector of measurable outputs. The system matrices $A \in \mathbb{R}^{n \times n}$, $B \in \mathbb{R}^{n \times N}$, $C \in \mathbb{R}^{(N+L) \times n}$, and $\Lambda \in \mathbb{R}^{N \times N}$ capture the aggregated dynamics of inertia and damping, SG compensation, IBR control, and inter-area power exchanges through AC tie-lines. The derivation of this model is presented in Section \ref{sec:SFR Model}.

\noindent \textbf{Problem 2:} Given the derived state-space model and a disturbance $\Delta P$ applied at $t=0$, design a control input for the IBRs that regulates system frequency to its nominal value in a power grid with high penetration of inverter-based renewable generation while satisfying operational constraints. The controller has access only to measurable outputs $y(t)$. The control design is addressed in Section \ref{sec:Optimal Control}.


\section{System Frequency  Response Model}\label{sec:SFR Model}

\begin{figure}[t]
\centering
\includegraphics[width=\columnwidth]{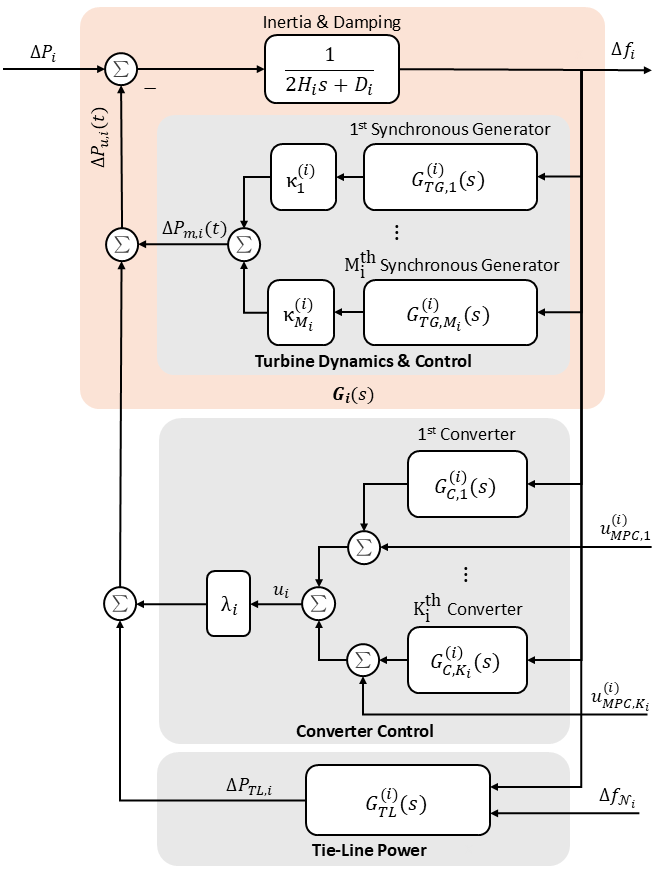}
\caption{SFR model for the \textit{i}-th operational region with $M_i$ SGs, $K_i$ converters, and tie-line power exchange.}
\label{fig:Single Area}
\end{figure}

In this section, we derive a linear state-space model for the SFR of an interconnected power grid. The model is constructed incrementally. We begin with inertia and damping dynamics of a single operational region, then augment the model to include SG governor compensation, inter-area coupling through tie-line dynamics, and finally IBR control, as illustrated in Fig. \ref{fig:Single Area}. At each step, the resulting dynamics are expressed in state-space form.

\subsection{Inertia and Damping Dynamics}\label{sec:SFR Inertia and Dampong}

In an operational region, maintaining a constant electrical frequency requires that, at any instant, the electrical power generated equals the power demanded by the load. Any imbalance between generation and demand results in deviations of the system frequency from its nominal value. These frequency dynamics can be modeled by a simple second-order differential equation, which provides a sufficiently accurate representation of how frequency evolves in response to power imbalances \cite{Anderson1990AModel}
\begin{equation}\label{eq:swing eq}
    2H_i \Delta \dot{f}_i + D_i \Delta f_i = \Delta P_i - \Delta P_{u,i}
\end{equation}
where $\Delta f_i(t)$ is the inertia-weighted average frequency deviation from the nominal value of the SGs in the $i$-th operational region,
which we will simply refer to as the regional frequency deviation moving forward. This regional frequency deviation
can be expressed using the aggregation \cite{Shi2018AnalyticalStudies}:
\begin{equation}\label{CoI f}
    \Delta f_i = \sum\nolimits_{m=1}^{M_i} \Delta f_m^{(i)} H_m^{(i)} \left/ \sum\nolimits_{m=1}^{M_i} H_m^{(i)} \right.
\end{equation}
where $\Delta f_m^{(i)}(t)$ and $H_m^{(i)}$ denote the frequency deviation and inertia constant of the \textit{m}-th SG in the \textit{i}-th operational region, respectively, and $M_i$ denotes the number of SGs in the \textit{i}-th operational region. The parameters $H_i$ and $D_i$ are the equivalent inertia constant and damping coefficient, respectively, given by \cite{Shi2018AnalyticalStudies}:
\begin{equation}\label{CoI H}
    H_i = \sum\nolimits_{m=1}^{M_i} S_m^{(i)} H_m^{(i)}\left/S_T^{(i)}\right.
\end{equation}
\begin{equation}\label{CoI D}
    D_i = \sum\nolimits_{m=1}^{M_i} S_m^{(i)} D_m^{(i)}\left/S_T^{(i)}\right.
\end{equation}
where $D_m^{(i)}$ and $S_m^{(i)}$ are the damping coefficient and rated apparent power of the \textit{m}-th SG in the \textit{i}-th operational region, respectively, and $S_T^{(i)}$ denotes the total rated apparent power of the \textit{i}-th operational region. Finally, $\Delta P_i(t)$ denotes the power disturbance in the \textit{i}-th operational region, while $\Delta P_{u,i}(t)$ represents the aggregate power compensation.

We now derive a state-space representation of (\ref{eq:swing eq}). We start by defining the regional frequency deviation as the state, $x_{f,i} \overset{\Delta}{=} \Delta f_i$, and the aggregate power compensation as
\begin{equation}\label{eq:u_bar}
\Delta P_{u,i} = \Delta P_{m,i} + \lambda_i u_i + \Delta P_{TL,i}
\end{equation}
where $\Delta P_{m,i}(t)$ is the primary control support from SGs, $\lambda_i u_i(t)$ represents the IBR support, and $\Delta P_{TL,i}(t)$ denotes the tie-line power flow to the \textit{i}-th operational region. Equation (\ref{eq:swing eq}) can then be written as
\begin{equation}\label{eq:x_f}
    \dot{x}_{f,i} = A_{f,i} x_{f,i} +
B_{f,i} \left( \Delta P_i - \Delta P_{m,i} - \lambda_i u_i - \Delta P_{TL,i} \right)
\end{equation}
\begin{equation}\label{eq:y_f}
y_{f,i} = x_{f,i}
\end{equation}
where $A_{f,i} = -D_i/2H_i$ and $B_{f,i} = 1/2H_i$.

The subsequent subsections explain how SGs, IBRs, and tie-line power contributions are incorporated into the model in (\ref{eq:x_f}) and (\ref{eq:y_f}) to obtain the state-space representation presented in (\ref{eq:state eq}) and (\ref{eq:meas eq}).

\subsection{Synchronous Generator Compensation}

In response to frequency deviations, SGs provide regulation through their turbine governor systems, which adjust the mechanical input power based on frequency measurements. Depending on the turbine type, several models have been developed, most notably \cite{Anderson1990AModel} for steam turbines, \cite{Rowen1983SimplifiedTurbines} for gas turbines, \cite{deMello1994DynamicStudies, Zhang2000DynamicStudies} for combined cycle gas turbines (CCGTs), and \cite{Machowski2020PowerControl} for hydro turbines. For small frequency deviations, typically within approximately 1 Hz, these turbine–governor models admit accurate linear representations \cite{Anderson1990AModel, Rowen1983SimplifiedTurbines, deMello1994DynamicStudies, Zhang2000DynamicStudies, Machowski2020PowerControl}. Let $x_{TG,i}(t)$ denote the internal states of all turbine governor dynamics in the \textit{i}-th region.
\begin{equation}
    x_{TG,i} = \begin{bmatrix}
    x_{TG,1}^{(i)T} &
    \cdots &
    x_{TG,M_i}^{(i)T}
    \end{bmatrix}^T
\end{equation}
where $x_{TG,m}^{(i)}(t)$ is the internal state of the \textit{m}-th turbine governor in the \textit{i}-th operational region. The aggregate turbine governor dynamics can be written as
\begin{equation}
    \dot{x}_{TG,i} = A_{TG,i}x_{TG,i} + B_{TG,i}x_{f,i}
\end{equation}
with the resulting mechanical power contribution
\begin{equation}\label{eq:u_SG}
    \Delta P_{m,i} = \sum\nolimits_{m=1}^{M_i}\kappa_{m}^{(i)}w_{TG,m}^{(i)}x_{TG,m}^{(i)}
\end{equation}
where $A_{TG,i}$, $B_{TG,i}$, and $w_{TG,m}^{(i)}$ are matrices of appropriate dimensions. The parameter $\kappa_m^{(i)}$ weights their relative contribution based on each SG’s rated apparent power and is defined as \cite{Shi2018AnalyticalStudies}
\begin{equation}\label{eq:kappa}
    \kappa_m^{(i)} \overset{\Delta}{=} S_m^{(i)}/S_T^{(i)}
\end{equation}

Substituting (\ref{eq:u_SG}) into (\ref{eq:x_f}) yields the augmented single-area state-space model
\begin{equation}\label{eq:state regional}
    \dot{x}_i = A^{(i)}x_{i} + B^{(i)}\left( \Delta P_i - \lambda_i u_i - \Delta P_{TL,i} \right)
\end{equation}
\begin{equation}\label{eq:meas regional}
    y_i = C^{(i)}x_i = x_{f,i}
\end{equation}
with the state vector
\begin{equation}\label{eq:single area state}
    x_i = \begin{bmatrix}
    x_{f,i} &
    x_{TG,i}^T
    \end{bmatrix}^T
\end{equation}

The single-area frequency response structure is illustrated in Fig. \ref{fig:Single Area}, where the inertia and damping dynamics are combined with turbine-governor compensation in the transfer function
\begin{equation}\label{eq:Gi}
    G_i(s) = C^{(i)}\left(sI-A^{(i)} \right)^{-1}B^{(i)}
\end{equation}
and the turbine governor dynamics for each SG are represented by the frequency-to-power transfer function $G_{TG,m}^{(i)}(s)$, rather than a state-space representation, for ease of notation.
We assume that the control parameters of the SGs are fixed.

\subsection{Interconnection through Tie-Line Dynamics}

In this section, we quantify the transmitted power $\Delta P_{TL,i}$ in (\ref{eq:x_f}). In interconnected power systems, frequency regulation within each operational region is further influenced by power exchanges with neighboring regions. These interactions are captured by the tie-line power flows, whose deviations from scheduled values reflect imbalances between regions and directly affect the local frequency dynamics. The net transmitted power deviation through AC tie-lines between the \textit{i}-th operational region and its interconnected neighbors $\mathcal{N}_i$ is given by
\begin{equation}\label{eq:TieLinePower}
    \Delta P_{TL,i} = \sum\nolimits_{j\in\mathcal{N}_i} \int 2\pi \frac{T_{ij}}{S_{T}^{(i)}}\left(\Delta f_i - \Delta f_j\right) dt
\end{equation}
where $T_{ij}$ denotes the synchronizing coefficient \cite{Alhelou2025DistributedSystem}.


The time-domain representation of the tie-line dynamics in (\ref{eq:TieLinePower}) is represented in frequency domain in Fig. \ref{fig:Single Area} in the form of a transfer function $G^{(i)}_{TL}(s)$, with
\begin{equation}
    \Delta f_{\mathcal{N}_i} = \left[\Delta f_{\mathcal{N}_i}^{(i)} \ \cdots \ \Delta f_{\mathcal{N}_i}^{(|\mathcal{N}_i|)} \right]^T
\end{equation}
denoting the vector of frequency deviations of the regions interconnected with the \textit{i}-th operational region. 

By combining the $N$ state-space models for each region given in (\ref{eq:state regional}) and (\ref{eq:meas regional}) with the definition of $\Delta P_{TL,l}$ in (\ref{eq:TieLinePower}), we can write the SFR model in the form (\ref{eq:state eq}), where the state vector $x(t)$ and output vector $y(t)$ are defined as
\begin{equation}\label{eq:state}
    x = \begin{bmatrix}
        x_1^T &
        \cdots &
        x_N^T &
        \Delta P_{TL,1}^T &
        \cdots &
        \Delta P_{TL,L}^T
    \end{bmatrix}^T
\end{equation}
\begin{equation}\label{eq:measurement}
    y = \begin{bmatrix}
        \Delta f_1 &
        \cdots &
        \Delta f_N &
        \Delta P_{TL,1} &
        \cdots &
        \Delta P_{TL,L}
    \end{bmatrix}^T
\end{equation}
with $x_i$ given by (\ref{eq:single area state}).
A representative set of states for the case study of the KSA power grid is shown in Appendix \ref{app:A}. 

The matrix $\Lambda$ in (\ref{eq:state eq}) is a diagonal matrix whose \textit{i}-th component $\lambda_i$ represent the share of renewable generation and is defined as \cite{Shi2018AnalyticalStudies}
\begin{equation}\label{eq:Renewable Share}
    \lambda_i \overset{\Delta}{=}   \sum\nolimits_{k=1}^{K_i} S_{IBR,k}^{(i)}\left/ S_T^{(i)}\right.
\end{equation}
where $S_{IBR,k}^{(i)}$ denotes the rated apparent power of the \textit{k}-th converter in the \textit{i}-th region and $K_i$ denotes the number of grid-forming converter in the \textit{i}-th region (Fig. \ref{fig:Single Area}). The control input $u(t)$ in the SFR model corresponds to the total grid IBR support, where the \textit{i}-th component $u_i(t)$ defined in (\ref{eq:u_bar}) represents the net IBR contribution from the $K_i$ converters in the \textit{i}-th operational region. The design of this control input is discussed in Section \ref{sec:Optimal Control}. 

So far, we have shown how to derive a detailed linear continuous-time state-space model starting from equation (\ref{eq:swing eq}) and incorporating the power contributions from SGs and tie-lines, thereby addressing Problem 1 of this paper. In the following section, we use representative data from the KSA power grid to illustrate the consequences of a high penetration of IBRs when they don't provide frequency response ancillary services, that is, when $u(t) = 0$.

\begin{table*}[b]
\centering
\caption{Pre-RES Energy Generation Mix in the Kingdom of Saudi Arabia}
\label{tab:Current Mix KSA}
\begin{tabular}{|c|c|c|c|c|c|}
\hline
\textbf{Type} & \textbf{Eastern {[}GW{]}} & \textbf{Central {[}GW{]}} & \textbf{Western {[}GW{]}} & \textbf{Southern {[}GW{]}} & \textbf{Total {[}GW{]}} \\ \hline
Steam         & $22.3$                    & $0.0^{\dagger}$             & $22.3$                    & $0.0^{\dagger}$              & $44.6^{\dagger}$                  \\ \hline
Gas/Diesel    & $6.2$                     & $12.7$                     & $6.3$                     & $6.7$                      & $31.9^{\dagger}$                  \\ \hline
CCGT          & $4.5$                     & $4.4$                     & $4.4$                     & $0.0^{\dagger}$              & $13.3^{\dagger}$                   \\ \hline
Renewable     & $1.3$           & $0.0$          & $0.0$          & $0.0$            & $1.3^{\dagger}$        \\ \hline
Total         & $34.3^{\dagger}$                    & $17.1^{\dagger}$                    & $33.0^{\dagger}$                    & $6.7^{\dagger}$                      & $91.1^{\dagger}$         \\ \hline
\multicolumn{6}{l}{$^{\dagger}$Values provided by \cite{SaudiElectricityRegulatoryAuthoritySERA2023AnnualElectricity}}
\end{tabular}
\end{table*}
\begin{table*}[b]
\centering
\caption{Projected 2030 Energy Generation Mix and Its Impact on Regional Inertia in the Kingdom of Saudi Arabia}
\label{tab:Inertia KSA}
\begin{tabular}{|c|c|c|c|c|c|}
\hline
\textbf{Type} & \textbf{Eastern {[}GW{]}} & \textbf{Central {[}GW{]}} & \textbf{Western {[}GW{]}} & \textbf{Southern {[}GW{]}} & \textbf{Total {[}GW{]}} \\ \hline
Steam         & $15.3$                    & $0.0^{\dagger}$             & $9.9$                    & $0.0^{\dagger}$              & $25.2$                  \\ \hline
Gas/Diesel    & $4.2$                     & $3.8$                     & $2.8$                     & $2.7$                      & $13.5$                  \\ \hline
CCGT          & $3.0$                     & $1.3$                     & $2.0$                     & $0.0^{\dagger}$              & $6.3$                   \\ \hline
Renewable     & $8.0^{\dagger}$           & $26.0^{\dagger}$          & $33.0^{\dagger}$          & $8.0^{\dagger}$            & $75.0^{\dagger}$        \\ \hline
Total         & $30.5$                    & $31.1$                    & $47.7$                    & $10.7$                      & $120^{\dagger}$         \\ \hline
\textbf{}     & \textbf{Eastern {[}s{]}}  & \textbf{Central {[}s{]}}  & \textbf{Western {[}s{]}}  & \textbf{Southern {[}s{]}}  & \textbf{}               \\ \hline
Inertia ($H_i$)       & $3.69$                    & $0.82$                    & $1.54$                    & $1.23$                     &                         \\ \hline
\multicolumn{6}{l}{$^{\dagger}$Values provided by \cite{SaudiElectricityRegulatoryAuthoritySERA2023AnnualElectricity}}
\end{tabular}
\end{table*}

\subsection{Numerical Results: Impact of Projected 2030 IBR Penetration in the Kingdom of Saudi Arabia}\label{sec:Sim No u}

The KSA is currently undergoing a transition of energy generation under the so-called \textit{Vision 2030}, which, among other objectives, aims to reduce carbon emissions by 278 million tonnes per year (mtpa) by 2030 \cite{SaudiInitiative}. As of 2023, the KSA power grid has a generation capacity of 90 GW, relying heavily on fossil-fuel-based generation (approximately 98\%) \cite{SaudiElectricityRegulatoryAuthoritySERA2023AnnualElectricity}, to meet a demand with a peak of around 74 GW \cite{FuturePowerExpo.2024TheFuture}. For 2030, the Kingdom aims to have a generation capacity of 120 GW, 75 of which will come from RESs 
\cite{SaudiElectricityRegulatoryAuthoritySERA2023AnnualElectricity}. This entails that 62.5\% of the generation capacity will come from RESs. This high penetration of RESs may compromise frequency resilience if there are no control mechanisms in place that allow IBRs to provide support in case of a disturbance.

The KSA power grid consists of four operating regions: Central, Eastern, Western, and Southern. The regions are interconnected by AC tie-lines, where the tie-line capacities are estimated to be 7.5 GW between the Central and Eastern regions, 0.85 GW between the Central and Southern regions, and 3.0 GW between the Central and Western regions as well as between the Southern and Western regions \cite{SaudiElectricityRegulatoryAuthoritySERA2023AnnualElectricity}. 
Four types of power plants operate in the system: steam, gas, diesel, and combined cycle. Steam plants are located only in the Eastern and Western regions; CCGT units operate in all regions except the Southern; diesel generation is absent in the Central; and gas plants are present in all regions \cite{SaudiElectricityRegulatoryAuthoritySERA2023AnnualElectricity}.

Table \ref{tab:Inertia KSA} has been derived under several assumptions.  Following \cite{Alshehri2025MaintainingSupport}, the inertia constant of the KSA power grid is taken as a conservative value of $H_i = 5$ for all synchronous units, within the reported pre-RES range of 4–5. 
Prior to renewable integration, steam generation is equally split between Eastern and Western regions, combined-cycle generation is equally shared among Eastern, Central, and Western regions based on 2023 data \cite{SaudiElectricityRegulatoryAuthoritySERA2023AnnualElectricity}, and gas generation is adjusted to meet regional totals. Anticipated synchronous plant retirements by 2030 are modeled as a uniform reduction in steam, gas, and combined-cycle output, preserving each region’s pre-RES share of national generation,
as shown in Table \ref{tab:Current Mix KSA}.
Since rated apparent powers are unavailable, generation capacity is used as a proxy, justified by the near-unity power factors of the plants, enabling quantification of equivalent inertia using (\ref{CoI H}). Finally, the damping coefficient is assumed to be $D_i = 1$ for all four operational regions. This assumption is examined in Section \ref{sec:Numerical Results}, where we show that the proposed control architecture is not sensitive to the choice of damping coefficient.

We use a standard metric, Under-Frequency Load Shedding (UFLS), which denotes the minimum allowable frequency, to evaluate the impact of renewables. In the context of the KSA power grid, the Saudi Arabian Grid Code specifies a limit of $58.8 \ \text{Hz}$ during continuous operation \cite{SaudiElectricityRegulatoryAuthoritySERA2024TheCode}, while some industries use a more conservative threshold of $59.85 \ \text{Hz}$. The latter is used to denote the UFLS limit in this paper.

\begin{figure}[ht]
\centering
\includegraphics[width=0.96\columnwidth]{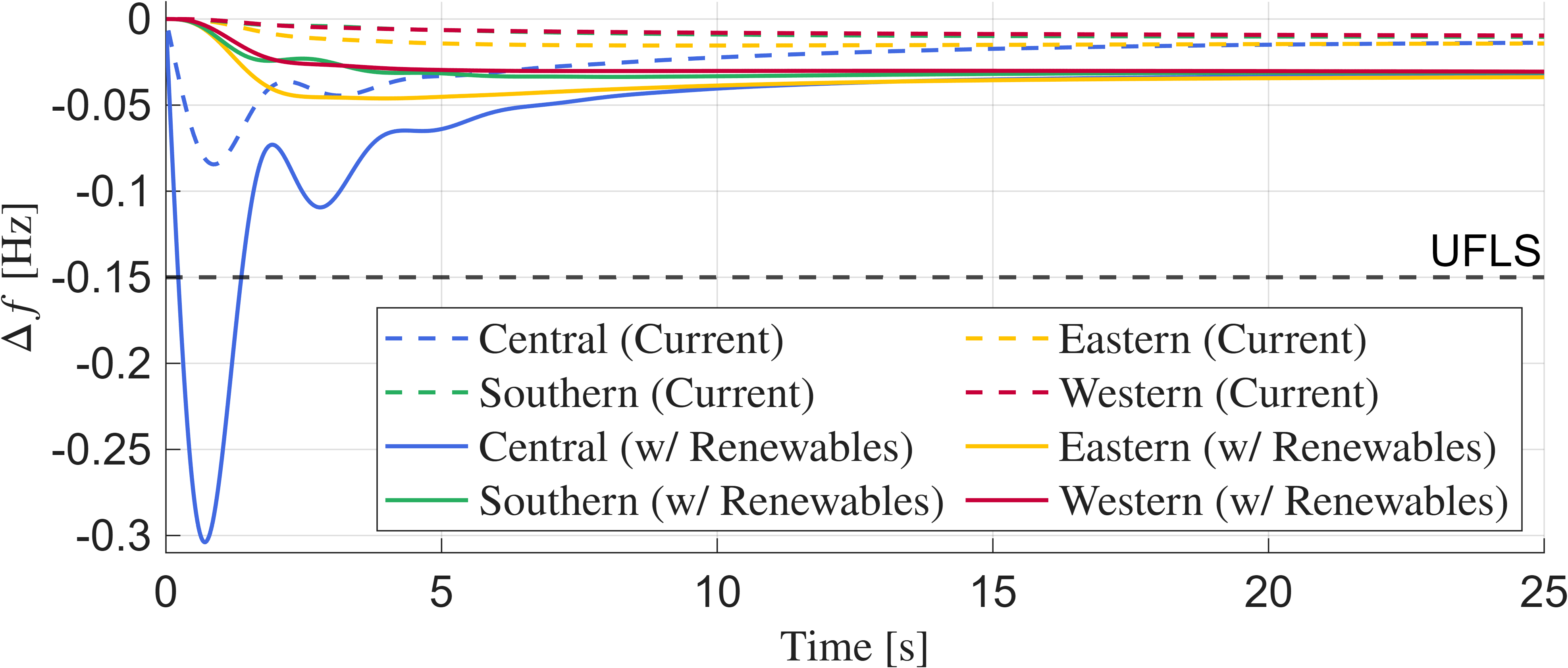}
\caption{Regional frequency deviations of the KSA grid following a $0.02\ \text{p.u.}$. disturbance in the Central region, shown before (dashed) and after (solid) the projected 2030 IBR penetration. In the latter case, IBRs do not provide ancillary services.}
\label{fig:RenewableNoControl}
\end{figure}

It should be noted  that Table \ref{tab:Inertia KSA}  represents the worst-case scenario based on installed generation for the 2030 energy mix by operational region, along with the corresponding expected system inertia under the planned integration of RES. With such a worst-case scenario, Table \ref{tab:Inertia KSA} indicates that the Central region is expected to experience the largest reduction in inertia, decreasing from $5$ to $0.82$ due to its projected highest share of renewable generation. Fig. \ref{fig:RenewableNoControl} illustrates the expected system frequency response when a $0.02 \  \text{p.u.}$ increase in load is applied to the Central region and IBRs do not provide ancillary services, that is, $u(t) = 0$. As can be observed, the high penetration of RES leads to a significant reduction in the frequency nadir, $\Delta f_{\text{nadir}}$, which crosses the UFLS.

To analyze the frequency dynamics in a tractable yet accurate manner, an aggregated representation of synchronous generation is adopted at the regional level. Although multiple SGs may exist within a single operational region, their collective dynamic response can be accurately represented by an aggregated single-machine (ASM) model \cite{Shi2018AnalyticalStudies}. The ASM parameters are obtained by computing an apparent-power-weighted average of the individual turbine governor parameters,
\begin{equation}\label{eq:ASM}
\chi^{(i)} = \sum_{m=1}^{M_i} \kappa_m^{(i)} \chi_m^{(i)}\left/ \sum_{m=1}^{M_i} \kappa_m^{(i)}\right.
\end{equation}
where $\chi_m^{(i)}$ denotes the parameter set of the \textit{m}-th SG in the \textit{i}-th operational region. This set includes parameters such as reheat, governor, and gas fuel time constants, among others.

For the simulations, up to three ASM models per region are considered, one for each type of synchronous generator present in the KSA: steam, gas/diesel, and CCGT. Steam turbines follow the classical formulation in \cite{Anderson1990AModel}, while gas turbines are represented by the linear model introduced in \cite{Rowen1983SimplifiedTurbines}. For CCGT units, steam turbines typically operate in sliding pressure mode and therefore do not respond significantly to governor action during the first seconds following a disturbance. Their contribution to power increase may take several minutes \cite{Pourbeik2003ModelingStudies}. Given the short simulation time horizon, the steam turbine dynamics of CCGT units are neglected, and only the gas turbine portion is modeled using the formulation in \cite{Rowen1983SimplifiedTurbines}. Typical parameter values reported in the literature are employed for all three ASM models. Following the aforementioned assumptions, generation capacities weight the contribution of each ASM model via the parameter $\kappa_m^{(i)}$ defined in (\ref{eq:kappa}). Appendix \ref{app:A} lists the state variables of the SFR model for the KSA grid.

The scenario shown in Fig. \ref{fig:RenewableNoControl} illustrates the consequences of a high penetration of inverter-based renewable generation in the absence of support from IBRs. These results stress the need for the contribution of such units, which is the focus of the following section.


\section{An Optimal Control Approach for Frequency Regulation in Multi-Area Power Grids}\label{sec:Optimal Control}

In this section, we present the design of the control input $u(t)$, addressing Problem 2 of this paper. The proposed control strategy employs a two-layer architecture combining primary and secondary control to provide both immediate frequency support and enforcement of operational constraints. The primary layer consists of droop and VSM controllers, whose parameters are optimized for each operational region to balance frequency deviations, RoCoF, and control effort. The secondary layer uses MPC to regulate the system frequency to its nominal value while ensuring constraint satisfaction.

The two-layer architecture proposed in this paper is essential, as each layer alone is inadequate. Droop and VSM controllers, being linear, cannot guarantee constraint compliance under all conditions, while secondary control may be too slow to provide this compliance. This is due to the fact that secondary control relies on SCADA systems that typically issue commands every 2 to 6 seconds. This update interval, denoted by $T_s$, will be accommodated in this paper using MPC in a judicious way so as to ensure that the overall architecture is fast enough to satisfy constraints and not cause triggering of protection mechanisms. Together, we will show that the two-layer architecture provides an optimal response with the primary controllers providing fast support, and the MPC layer determining complementary actions that enable optimization and constraint satisfaction.

\begin{figure}[h]
\centering
\includegraphics[width=0.98\columnwidth]{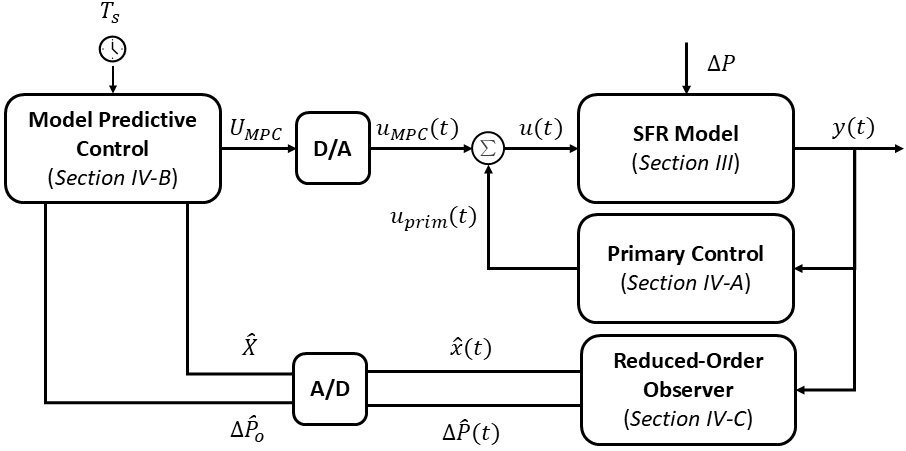}
\caption{Schematic of the proposed optimal control scheme for frequency regulation in multi-area power grids. A/D and D/A denote Analog-to-Digital and Digital-to-Analog converters, respectively.}
\label{fig:Block Diagram}
\end{figure}

Fig. \ref{fig:Block Diagram} illustrates the proposed control architecture, comprising four main components:
\begin{enumerate}
    \item \textbf{SFR model} (Section \ref{sec:SFR Model}): receives the total control input $u(t)=u_{prim}(t)+u_{MPC}(t)$ and the disturbance $\Delta P$, producing measured signals $y(t)$.
    \item \textbf{Primary control} (Section \ref{sec:Primary Control}): generates $u_{prim}(t)$ from $y(t)$ at a high update rate, thereby approximating continuous-time operation.
    \item \textbf{Reduced-order observer} (Section \ref{sec:Observer}): because $\Delta P$ is generally unknown, this block uses $y(t)$ and an initial disturbance estimate to produce estimates of the reduced-order state $\hat{x}(t)$ and an improved disturbance estimate $\Delta \hat{P}(t)$. After sampling through the A/D block, the resulting discrete signals $\hat{X}$ and $\Delta \hat{P}_o$ are fed to the MPC.
    \item \textbf{MPC} (Section \ref{sec:MPC}): solves a discrete-time optimization every $T_s$ seconds to enforce operational constraints, producing the control sequence $U_{MPC}$, which is converted to continuous time $u_{MPC}(t)$ via a zero-order hold.
\end{enumerate}

This structure ensures a rapid primary-layer response to disturbances, while the secondary layer restores nominal frequency and enforces grid constraints. To enable this regulation, we assume that sufficient power reserves exist in the grid.

\subsection{Nominal Controller for Primary Frequency Response}\label{sec:Primary Control}

Let $K_i$ be the number of converters that contribute to primary frequency control in the \textit{i}-th operational region through grid-forming control strategies. The frequency-to-power dynamics of the \textit{k}-th converter are represented in the frequency domain in Fig. \ref{fig:Single Area} by the transfer function $G_{c,k}^{(i)}(s)$, which may correspond to either droop (\ref{eq:G Droop}) or VSM control (\ref{eq:G VSM}) \cite{Markovic2019LQR-BasedPenetration}
\begin{equation}\label{eq:G Droop}
G_{\mathrm{Droop}}(s) = K_d \left/ \left(R_d(1 + T_d s)\right)\right.
\end{equation}
\begin{equation}\label{eq:G VSM}
G_{\mathrm{VSM}}(s) = \left(2H_c s + D_c\right) \left/ \left(1 + T_c s\right) \right.
\end{equation}
such that $T_d$ and $T_c$ denote the converter time constants associated with droop and VSM control, respectively. The parameters $R_d$ and $K_d$ represent the droop coefficient and power gain of droop-controlled converters, while $H_c$ and $D_c$ denote the normalized virtual inertia and damping coefficients of VSM-controlled converters, respectively.

To simplify the system model, converters are represented using an ASM model in which the converter parameters are averaged based on their rated apparent power within the \textit{i}-th operational region, analogous to (\ref{eq:ASM}). For droop-controlled converters, the ASM parameter set is defined as $\chi^{(i)} = \bigl\{K_{d}^{(i)}, R_{d}^{(i)}, T_d^{(i)}\bigr\}$, whereas for VSM-controlled converters it is $\chi^{(i)} = \bigl\{H_{c}^{(i)}, D_{c}^{(i)}, T_c^{(i)}\bigr\}$. Based on this ASM representation, the parameters of the \textit{i}-th operational region are then optimized by solving the following optimization problem:

\begin{mini!}
{H_c^{(i)}, D_c^{(i)}, \bar{K}_d^{(i)}, \gamma^{(i)}}{\frac{1}{T}\int_{0}^T J(t) \ dt \label{eq:cost function primary}}
{}{}
\addConstraint{H_{min} \leq H_c^{(i)} \leq H_{max}} \label{eq:Primary H}
\addConstraint{D_{min} \leq D_c^{(i)} \leq D_{max}} \label{eq:Primary D}
\addConstraint{\bar{K}_{min} \leq \bar{K}_d^{(i)} \leq \bar{K}_{max}} \label{eq:Primary K}
\addConstraint{0 \leq \gamma^{(i)} \leq 1}\label{eq:Primary gamma}
\end{mini!}
where $T$ denotes the optimization time horizon and $\bar{K}_d^{(i)} = K_d^{(i)}/R_d^{(i)}$. The constants $H_{min}$, $D_{min}$, $\bar{K}_{min}$ and $H_{max}$, $D_{max}$, $\bar{K}_{max}$ define the lower and upper bounds on $H_c^{(i)}$, $D_c^{(i)}$, and $\bar{K}_d^{(i)}$, respectively. Representative values for these bounds for the Australian National Electricity Market (NEM) are provided in \cite{Alhelou2025DistributedSystem}. The parameter $\gamma^{(i)}$ represents the proportion of droop- versus VSM-controlled converters in the \textit{i}-th operational region, where $\gamma^{(i)} = 0$ corresponds to all droop-controlled converters and $\gamma^{(i)} = 1$ corresponds to all VSM-controlled converters. The cost function $J(t)$ is defined as
\begin{equation}\label{eq:cost}
J = \mu_1^{(i)} \Delta f_i^2 + \mu_2^{(i)} \Delta \dot{f}_i^2 + \mu_3^{(i)} u_{\mathrm{prim},i}^2
\end{equation}
where $\mu_1^{(i)},\mu_2^{(i)},\mu_3^{(i)} > 0$ are weighting coefficients that penalize regional frequency deviations $\Delta f_i(t)$, RoCoF $\Delta \dot{f}i(t)$, and the primary control effort $u_{\mathrm{prim},i}(t)$, respectively. These signals are computed by neglecting tie-line power exchanges, enabling an independent control design for each region. 

The converter time constants $T_d^{(i)}$ and $T_c^{(i)}$ are excluded from the optimization, as they are fixed by the converters’ physical characteristics and are not readily adjustable \cite{Alhelou2025DistributedSystem}.

\noindent \textbf{Remark 1:} The optimization problem in (\ref{eq:cost function primary})-(\ref{eq:Primary gamma}) is nonlinear and non-convex, since the cost function (\ref{eq:cost}) depends nonlinearly on the decision variables through the closed-loop dynamics. The problem can be solved using a gradient-based nonlinear programming solver (e.g., MATLAB's \texttt{fmincon}, interior-point algorithm). To alleviate any sensitivity to local minima, the solver is run using twenty randomized initial guesses drawn uniformly within the bounds (\ref{eq:Primary H})-(\ref{eq:Primary gamma}), and only the best feasible solution across all runs is retained. Since this procedure is performed entirely offline during controller design, it introduces no computational burden during real-time operation of the proposed MPC framework.

While primary frequency control via droop and VSM can mitigate large frequency excursions, its linear nature prevents it from guaranteeing operational constraint satisfaction under all conditions and from fully restoring the frequency to its nominal value. These limitations motivate the use of secondary frequency control, which addresses both shortcomings and is the focus of the following section.

\subsection{Model Predictive Control for Secondary Frequency Control}\label{sec:MPC}

The secondary control layer has two main objectives: it coordinates resources across different operational regions to restore system frequency to its nominal value, and it enforces operational constraints. To achieve this, an MPC-based optimization is formulated over a prediction horizon of length $\Omega$, aiming to minimize both the control effort required from the IBRs and the frequency deviations across all operational regions, as expressed in (\ref{eq:MPC Cost Function})–(\ref{eq:MPC relaxation}).
\begin{figure*}[t!]
    \begin{mini!}
      {\Delta u_{1:\Omega-1}, \varepsilon_u, \varepsilon_f}{\eta_f^T\varepsilon_f + \eta_u^T\varepsilon_u + \sum_{k=1}^{\Omega-1} ||\hat{C}_1\hat{x}_k||^2_Q + ||\Delta u_k||^2_R\label{eq:MPC Cost Function}}
      {}{}
      \addConstraint{\hat{x}_{k+1} = \hat{A}_d\hat{x}_k + \hat{B}_d\left(\Delta \hat{P}_o - \Lambda \mu_k \right)}{,\quad}{\forall k \in \{0, \cdots, \Omega-1\}}\label{eq:MPC dynamics}
      \addConstraint{u_k = u_{k-1} + \Delta u_k}{,\quad}{\forall k \in \{1, \cdots, \Omega-1\}}\label{eq:MPC control update}
      \addConstraint{\mu_k = u_k + \nu_k}{,\quad}{\forall k \in \{1, \cdots, \Omega-1\}}\label{eq:MPC plus prim}
      \addConstraint{\underline{\Delta f} - \varepsilon_f \preceq f_0\hat{C}_1\hat{x}_k \preceq \overline{\Delta f} + \varepsilon_f}{,\quad}{\forall k \in \{1, \cdots, \Omega\}}\label{eq:MPC frequency}
      \addConstraint{\left|\left|\frac{f_0}{n_r\Delta t}\hat{C}_1\left( \hat{x}_{n_r+k} - \hat{x}_k \right) \right|\right|_{\infty} \leq M}{,\quad}{\forall k \in \{0, \cdots, \Omega-n_r\}}\label{eq:MPC rocof}
      \addConstraint{\underline{P}_{TL} \preceq \hat{C}_2\hat{x}_k \preceq \overline{P}_{TL}}{,\quad}{\forall k \in \{1, \cdots, \Omega\}}\label{eq:MPC tie line}
      \addConstraint{\left|\left|\Delta u_k \right|\right|_{\infty} \leq \overline{\Delta u}}{,\quad}{\forall k \in \{1, \cdots, \Omega-1\}}\label{eq:MPC slew rate}
      \addConstraint{\left(\mathbf{0} - \Bar{P}^{\star}_{IBR} \right) \preceq \mu_k \preceq \left(\mathbf{1} - \Bar{P}^{\star}_{IBR} \right) }{,\quad}{\forall k \in \{1, \cdots, \Omega-1\}}\label{eq:MPC headroom}
      \addConstraint{\underline{u} - \varepsilon_u \preceq \mu_k \preceq \overline{u} + \varepsilon_u}{,\quad}{\forall k \in \{1, \cdots, \Omega-1\}}\label{eq:MPC power}
      \addConstraint{\varepsilon_u, \varepsilon_f \succeq 0}{\quad}{}\label{eq:MPC relaxation}
    \end{mini!}
\end{figure*}

Here, $\hat{x}_k \in \mathbb{R}^r$ denotes the estimated state vector at discrete time step $k$, where $r$ is the dimension of the reduced-order model (see Section \ref{sec:Observer}). The variable $\Delta u_k \in \mathbb{R}^N$ represents the change in converter active power setpoints between time steps $k$ and $k-1$. The matrix $\hat{C}_1 \in \mathbb{R}^{N \times r}$ maps the reduced-order state to the $N$ regional frequency deviations, as described in Section \ref{sec:Observer}. The symmetric positive definite matrices $Q, R \in \mathbb{S}^{N \times N}_{++}$ in the cost function (\ref{eq:MPC Cost Function}) penalize the regional frequency deviations and the setpoint increments, respectively, and $||x||_A^2$ denotes $x^TAx$. Penalizing $\Delta u_k$ rather than the absolute setpoints $u_k$ is standard when regulation requires a non-zero steady-state control effort, as is the case for secondary frequency control.

The constraints of the optimization problem are given in (\ref{eq:MPC dynamics})-(\ref{eq:MPC relaxation}). The prediction model, defined in (\ref{eq:MPC dynamics}) and explained in Section \ref{sec:Observer}, allows the MPC to anticipate the power grid’s behavior. Constraint (\ref{eq:MPC control update}) defines the relationship between the absolute setpoint change $u_k$ and the decision variables $\Delta u_k$. Relation (\ref{eq:MPC plus prim}) represents the combined action of the primary control input $\nu_k \in \mathbb{R}^N$ and the secondary control input $u_k$. The \textit{i}-th entry of $\nu_k$ is obtained by applying Tustin's method
\begin{equation}\label{eq:Tustin}
    s \approx \frac{2}{\Delta t}\frac{z-1}{z+1},
\end{equation}
to the optimized primary controller transfer function of the \textit{i}-th operational region:
\begin{equation}\label{eq:primaryTF}
F_i(s;\theta^{\star}) = \left(1 - \gamma^{(i)\star}\right)\frac{\bar{K}_d^{(i)\star}}{1 + T_d^{(i)} s}
+ \gamma^{(i)\star}\frac{2H_c^{(i)\star} s + D_c^{(i)\star}}{1 + T_c^{(i)} s}
\end{equation}
where $z$ is the complex frequency variable of the $z$-transform, and $\theta^{\star} = \bigl\{\bar{K}_{d}^{(i)\star}, H_{c}^{(i)\star}, D_{c}^{(i)\star}, \gamma^{(i)\star}\bigr\}$ are the optimized droop and VSM parameters. The resulting discrete-time filter $\tilde{F}_i(z;\theta^{\star})$ is obtained by substituting (\ref{eq:Tustin}) into (\ref{eq:primaryTF}). This filter is then embedded in the MPC prediction model: at each step of the prediction horizon, the predicted frequency deviation is propagated through $\tilde{F}_i(z;\theta^\star)$ to produce the predicted primary control input $\nu_k$, which in turn enters the state update. The internal states of $\tilde{F}_i(z;\theta^\star)$ are therefore included in the augmented state vector used for MPC prediction.

Constraint (\ref{eq:MPC frequency}) enforces regional frequency limits. Here, $\underline{\Delta f}$ and $\overline{\Delta f}$ denote the lower and upper bounds of the allowable frequency band, and $f_0$ is the nominal frequency. A slack variable $\varepsilon_f \succeq 0$ is introduced to preserve feasibility, which is necessary because the MPC updates control inputs only every $T_s$ seconds, and disturbances may occur for which primary control cannot guarantee that frequency remains within the specified bounds. The slack variable $\varepsilon_f$ is heavily penalized in the cost function (\ref{eq:MPC Cost Function}) through the weighting vector $\eta_f \in \mathbb{R}^N$, encouraging the frequency to return rapidly to the admissible band. Constraint (\ref{eq:MPC rocof}) enforces a bound on the average RoCoF, ensuring it does not exceed the maximum allowable value $M$. The average is computed over a time window of length $n_r \Delta t$, where $\Delta t$ is the MPC time step and $n_r < \Omega$. Average RoCoF constraints are preferred over instantaneous measurements to filter out noise and to trigger protective actions only when real, sustained frequency emergencies occur. For example, the KSA's \cite{SaudiElectricityRegulatoryAuthoritySERA2024TheCode} and Great Britain's \cite{NationalEnergySystemOperator2026The41} grid codes establish a measuring window of $500$ milliseconds, while Australia's \cite{AustralianEnergyMarketCommissionAEMC2023FrequencyStandard} specifies windows between $250$ and $500$ milliseconds depending on the type of contingency and whether it occurs on the mainland or in Tasmania. 

Constraint (\ref{eq:MPC tie line}) limits tie-line power flows. The matrix $\hat{C}_2 \in \mathbb{R}^{L \times r}$ maps the reduced-order model state to the $L$ tie-line power flows, as detailed in Section \ref{sec:Observer}, while $\underline{P}_{TL}$ and $\overline{P}_{TL}$ define their lower and upper bounds. Constraint (\ref{eq:MPC slew rate}) restricts the control slew rate, with $\overline{\Delta u}$ representing the maximum allowable setpoint change over $\Delta t$ to prevent inverter tripping.

Constraints (\ref{eq:MPC headroom}) and (\ref{eq:MPC power}) represent operational limits associated with available IBR headroom and scheduled power reserves, respectively. The quantity $\bar{P}_{IBR}^{\star}$ denotes the operating active power setpoint of the IBRs, while $\underline{u}$ and $\overline{u}$ define the lower and upper bounds on scheduled power outputs, which may result from day-ahead commitments. The headroom constraint reflects a physical limitation on the maximum deliverable power, whereas the scheduled power constraint is typically more restrictive and reflects economic considerations. Constraint (\ref{eq:MPC power}) can be relaxed using a slack variable $\varepsilon_u \succeq 0$ when additional power is required to compensate for disturbances beyond the scheduled allocation. Similar to $\varepsilon_f$, $\varepsilon_u$ is heavily penalized in the cost function (\ref{eq:MPC Cost Function}) via the weighting vector $\eta_u \in \mathbb{R}^N$, encouraging the control action to remain within the scheduled power limits. Finally, constraint (\ref{eq:MPC relaxation}) enforces the non-negativity of the slack variables $\varepsilon_f$ and $\varepsilon_u$.

Once the MPC problem is solved, control commands corresponding to the next $T_s$ seconds are transmitted to the IBRs while the subsequent MPC optimization is computed. This process is repeated every $T_s$ seconds. We denote the resulting sequence of control actions as
\begin{equation}
U_{MPC} = \begin{bmatrix}
u_1^{\star} & u_2^{\star} & \cdots & u_h^{\star}
\end{bmatrix}
\end{equation}
as illustrated in Fig. \ref{fig:Block Diagram}, where $u_k^{\star}$ are the optimal setpoint changes, $h < \Omega$, and $h \Delta t = T_s$. The condition $h < \Omega$ ensures that the MPC optimizes over a prediction horizon longer than the interval during which control inputs are actually applied. This allows the controller to anticipate longer-term system behavior, avoiding myopic decisions and improving overall performance.

\begin{figure}[h]
\centering
\includegraphics[width=0.99\columnwidth]{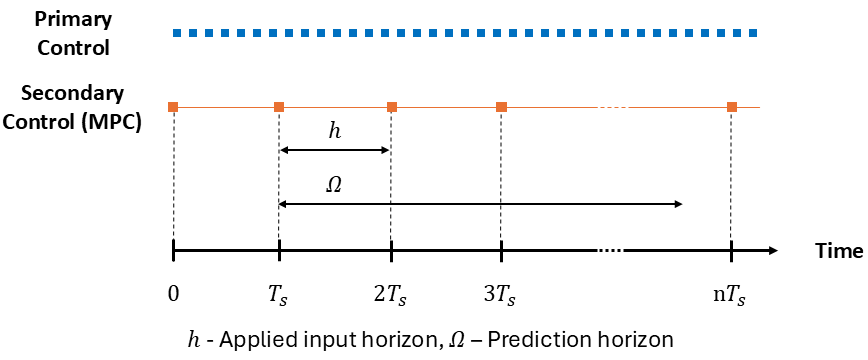}
\caption{Timing coordination between primary and secondary control. The primary runs at a high update rate, approximating continuous operation (top, blue squares), while the secondary MPC layer issues control commands at discrete intervals of $T_s$ seconds (bottom, orange squares).}
\label{fig:Timing}
\end{figure}

Fig. \ref{fig:Timing} illustrates the overall timing coordination: the primary control layer acts at a high update rate, while the secondary MPC layer issues commands only every $T_s$ seconds. At each activation, the MPC solves the optimization problem over the full prediction horizon $\Omega$ but applies only the first $h < \Omega$ steps before recomputing at the next activation, ensuring that control decisions account for system behavior beyond the immediately applied interval.

The control sequence $U_{MPC}$ is converted into a continuous-time signal $u_{MPC}(t)$ via a zero-order hold. Its \textit{i}-th component $u_{MPC}^{(i)}(t)$ corresponds to the sum of the $K_i$ converter setpoints in the \textit{i}-th operational region denoted as $u_{MPC,k}^{(i)}(t)$ in Fig. \ref{fig:Single Area}.

We conclude this section with a discussion of stability, recursive feasibility, and constraint satisfaction under model mismatch.

Since the inertia and synchronous generator turbine-governor dynamics that underlie the power grid model provide an inherently stabilizing compensatory response to frequency deviations, the resulting power grid model is inherently stable. Combined with the quadratic MPC cost, closed-loop stability follows from the results in \cite{Mayne2000ConstrainedOptimality}. These guarantees can be made rigorous by augmenting the optimization formulation with a terminal cost equal to the LQR cost-to-go and a suitably constructed terminal set, as described in \cite{Mayne2000ConstrainedOptimality}. The simulations of Section \ref{sec:Numerical Results} confirm that stability and feasibility are maintained without these elements, and we have therefore omitted them to keep the formulation concise; the theoretical tools are nevertheless available should stricter guarantees be required.

Performance constraints are softened via slack variables with a large penalty weight, guaranteeing that the QP is always solvable provided physical generation constraints can be met. Physical limits on generation are retained as hard constraints; if the total available reserve is insufficient to compensate for a given disturbance, the MPC delivers best-effort performance within the available resources.

Finally, since both the reduced-order state and the disturbance are jointly detectable with the proposed observer (Section \ref{sec:Observer}), the estimation and model mismatch errors are bounded and enter the closed-loop as a bounded additive disturbance, which the slack-variable formulation absorbs while the large penalty weight ensures constraint enforcement is maintained in practice.

\noindent \textbf{Remark 2:} We note that the proposed MPC replaces the conventional PI-based AGC controller, rather than augmenting it. This allows the framework to explicitly enforce operational constraints while respecting the control-room update rates of the existing primary-secondary control infrastructure.

\noindent \textbf{Remark 3:} Deadbands are commonly used in practical AGC implementations: once the frequency deviation falls within this band, the primary regulation contribution is deactivated, which can cause frequency re-deterioration and power and frequency fluctuations. The proposed architecture can accommodate this behavior through supervisory logic: as the frequency deviation approaches the deadband, a command can deactivate primary regulation and transfer the final recovery stage to the MPC, since the role of primary control is mainly to arrest the initial frequency excursion. A detailed treatment of this supervisory logic is left for future work, as it lies outside the scope of the reduced-order modeling and observer-based MPC architecture presented here.

\subsection{Reduced-Order Observer}\label{sec:Observer}

Since the MPC optimization requires the current state of the SFR model to predict power grid behavior, and this state is not directly accessible, we formulate a reduced-order observer to estimate the state. To develop the reduced-order observer, we first perform a balanced truncation based on the HSVs of the SFR model. Subsequently, we employ a Kalman-Bucy filter to estimate the reduced-order state and disturbance estimate. This approach provides a quantifiable model reduction that does not rely on heuristics and is not restricted to specific types of turbine models.

\subsubsection{Balanced Truncation}

Starting from the transfer function $G_i(s)$ in (\ref{eq:Gi}), a minimal representation is derived, which we denote by $\bar{G}_i(s)$. From this minimal single-area dynamic model, a balanced state-space realization is obtained, which we denote as
\begin{equation}
    \Xi_i = \left(
        \bar{A}^{(i)}, \bar{B}^{(i)},
        \bar{C}^{(i)}, \bar{D}^{(i)} =0
    \right)
\end{equation}
where $\bar{A}^{(i)} \in \mathbb{R}^{n_i \times n_i}$, $\bar{B}^{(i)} \in \mathbb{R}^{n_i}$, $\bar{C}^{(i)} \in \mathbb{R}^{1 \times n_i}$, and $n_i$ is the state dimension of the \textit{i}-th operational region.

Several observations are in order:
\begin{enumerate}
    \item A balanced state-space realization is one in which the controllability Gramian, $W_c$, and the observability Gramian, $W_o$, are equal and diagonal, that is,
    \begin{equation}
        W_c = W_o = \text{diag}(\sigma_1, \ \cdots, \ \sigma_{n_i})
    \end{equation}
    where $\sigma_i$ are the HSVs and satisfy $\sigma_{1} \geq \cdots \geq \sigma_{n_i} > 0$ \cite{Antoulas2004ApproximationOverview}. The HSVs provide a measure of the energy associated with each state of the system.
    \item The use of a minimal representation of $G_i(s)$ guarantees that $\Xi_i$ is both controllable and observable, which is necessary for performing the balanced truncation \cite{Antoulas2004ApproximationOverview}.
\end{enumerate}

By partitioning the matrices $\bar{A}^{(i)}$, $\bar{B}^{(i)}$, and $\bar{C}^{(i)}$ as (\ref{eq:partition}), the system (\ref{eq:redux system}) is an $r_i$ reduced-order system obtained from $\Xi_i$ via balanced truncation, where $\bar{A}_{11}^{(i)} \in \mathbb{R}^{r_i \times r_i}$, $\bar{B}_{1}^{(i)} \in \mathbb{R}^{r_i}$, and $\bar{C}_{1}^{(i)} \in \mathbb{R}^{1 \times r_i}$, with $r_i < n_i$.
\begin{equation}\label{eq:partition}
    \bar{A}^{(i)} = \begin{bmatrix}
        \bar{A}_{11}^{(i)} & \bar{A}_{12}^{(i)} \\
        \bar{A}_{21}^{(i)} & \bar{A}_{22}^{(i)}
    \end{bmatrix}
\end{equation}
\begin{equation}
    \bar{B}^{(i)} = \begin{bmatrix}
        \bar{B}_{1}^{(i)} \\
        \bar{B}_{2}^{(i)} 
    \end{bmatrix}
\end{equation}
\begin{equation}
    \bar{C}^{(i)} = \begin{bmatrix}
        \bar{C}_{1}^{(i)} & \bar{C}_{2}^{(i)}
    \end{bmatrix}
\end{equation}
\begin{equation}\label{eq:redux system}
    \hat{\Xi}_i = \begin{bmatrix}
       \bar{A}_{11}^{(i)} & \bar{B}_{1}^{(i)} \\
       \bar{C}_{1}^{(i)} & 0
    \end{bmatrix}
\end{equation}

The following properties hold for the balanced truncation
$\hat{\Xi}_i$ \cite{Antoulas2004ApproximationOverview}:
\begin{enumerate}
    \item $\hat{\Xi}_i$ is balanced, controllable, and observable.
    \item The $\mathcal{H}_{\infty}$-norm of the 
    error between $\Xi_i$ and $\hat{\Xi}_i$ is bounded by twice the sum of the neglected HSVs, i.e.,
    \begin{equation}
        \left|\left|\Xi_i-\hat{\Xi}_i\right|\right|_{\mathcal{H}_{\infty}}\leq 2\sum\nolimits_{i=r_i+1}^{n_i}\sigma_i
    \end{equation}
\end{enumerate}

This result implies that the choice of $r_i$ directly determines the desired accuracy of the reduced-order model for the \textit{i}-th operational region.

Once the reduced-order systems for each operational region have been obtained, they are assembled into a single state-space model as explained in Section \ref{sec:SFR Model}, to account for the tie-line power flows. This results in the following reduced-order state-space model:
\begin{align}
    \dot{\hat{x}}(t) & = \hat{A}\hat{x}(t) + \hat{B}\left(\Delta \hat{P}(t) - \Lambda u(t)\right) \label{eq:reduced order x} \\
    \hat{y}(t) & = \hat{C}\hat{x}(t) \label{eq:reduced order y}
\end{align}
with $\hat{x} \in \mathbb{R}^r$, $\hat{A} \in \mathbb{R}^{r \times r}$, $\hat{B} \in \mathbb{R}^{r \times N}$, $\hat{C} \in \mathbb{R}^{(N+L) \times r}$, and $r = L + \sum_{i=1}^N r_i$. Without loss of generality, the first $N$ outputs are assumed to correspond to the regional frequency deviations, while the remaining $L$ outputs correspond to tie-line power flows. Accordingly, the output matrix is partitioned as
\begin{equation}
    \hat{C} = \begin{bmatrix}
        \hat{C}_1^T &
        \hat{C}_2^T 
    \end{bmatrix}^T
\end{equation}
where $\hat{C}_1 \in \mathbb{R}^{N \times r}$ maps the reduced-order state to regional frequency deviations and $\hat{C}_2 \in \mathbb{R}^{L \times r}$ maps the reduced-order state to tie-line power flows, as referenced in Section \ref{sec:MPC}.


To obtain the discrete-time model used in the MPC formulation, a zero-order hold on the inputs is assumed. Under this assumption, the exact discrete-time system and input matrices are given by
\begin{equation}
    \hat{A}_d = e^{\Delta t \hat{A}}
\end{equation}
\begin{equation}
    \hat{B}_d = \left(\int_{0}^{\Delta t} e^{\tau \hat{A}} \ d \tau\right)\hat{B}
\end{equation}

\noindent \textbf{Remark 4:} Although balanced truncation could, in principle, be applied directly to the full power grid model, this modular approach is adopted because, for large-scale systems, computing a balanced realization may be numerically unstable \cite{Antoulas2004ApproximationOverview}. Moreover, it allows reduced-order models of individual regions to be updated independently, without requiring recomputation of a reduced-order model for the entire system.

\subsubsection{Observer}

We leverage a Kalman–Bucy filter to estimate online the state of the reduced-order model $\hat{x}$ as well as the disturbance $\Delta \hat{P}$. The filter dynamics are given by
\begin{multline}
    \begin{bmatrix}
        \dot{\hat{x}} \\
        \Delta \dot{\hat{P}}
    \end{bmatrix} = \underbrace{\begin{bmatrix}
        \hat{A} & \hat{B} \\
        0 & 0
    \end{bmatrix}}_{\bar{A}} \begin{bmatrix}
        \hat{x} \\
        \Delta \hat{P}
    \end{bmatrix} - \begin{bmatrix}
        \hat{B} \\ 0
    \end{bmatrix} \Lambda u -
    M\left( \hat{C}\hat{x} - y \right)
\end{multline}
where $M \in \mathbb{R}^{(r+N) \times (N+L)}$ is given by
\begin{equation}
    M^T = R_{\nu}^{-1}\underbrace{\begin{bmatrix}
        \hat{C} & 0
    \end{bmatrix}}_{\bar{C}}\Sigma
\end{equation}
and $\Sigma \in \mathbb{R}^{(r+N) \times (r+N)}$ is the solution to the algebraic Riccati equation
\begin{equation}
    \bar{A}\Sigma + \Sigma\bar{A}^T - \Sigma\bar{C}^TR_{\nu}^{-1}\bar{C}\Sigma + Q_{\omega} = 0
\end{equation}
such that $Q_{\omega} \in \mathbb{S}_{+}^{r+N}$ and $R_{\nu} \in \mathbb{S}_{++}^{N+L}$ are tuning hyperparameters.

The disturbance-augmented system is detectable for general power grids; we provide sufficient conditions for this claim in Appendix \ref{appendix A}, which are generically satisfied in practice. We also highlight that our approach does not require direct state observation, as all constraints are defined in terms of the system outputs: frequency deviations and tie-line power flows. The reduced-order model provides a representation that accurately predicts the evolution of these outputs. Since both the reduced-order state and disturbance are jointly detectable, the estimation error converges to zero in the absence of noise and remains bounded by the noise covariance in the stochastic setting. The slack variable formulation in the MPC absorbs this error, preserving feasibility, while the heavy penalty on constraint violations ensures that enforcement is maintained in practice.

Using a filter to estimate a constant, such as the disturbance $\Delta \hat{P}(t)$, is a practice inspired by applications including Simultaneous Localization and Mapping (SLAM) \cite{Dissanayake2001AProblem}. To the best of our knowledge, this is the first application of a Kalman–Bucy filter for estimating the state of an interconnected power grid.

\noindent \textbf{Remark 5:} The matrices $Q_\omega$ and $R_\nu$ determine the estimation properties of the observer. Larger values $R_\nu$ relative to $Q_\omega$ reflect less confidence in the measurements, while larger values of $Q_\omega$ relative to $R_\nu$ reflect less confidence in the accuracy of the model. In practice, these matrices are often chosen as diagonal, where the $i$-th diagonal entry corresponds to the $i$-th operational region, allowing one to capture the trade-off between model and measurement accuracy individually for each region.

In this section, we have presented a two-layer control architecture for frequency regulation in grids with a high penetration of IBR generation. The proposed architecture balances performance at both the primary and secondary levels while handling constraints. At the primary level, controller gains are constrained to reflect converter limits, whereas at the secondary level, operational constraints are explicitly incorporated into the MPC optimization. The control architecture operates at realistic control update rates and relies only on measurable outputs, thereby addressing Problem 2 of this paper. In the following section, representative data from the KSA power grid is used to demonstrate the effectiveness of the proposed approach.


\section{Numerical Results}\label{sec:Numerical Results}

Section \ref{sec:Sim No u} showed the consequences of large-scale integration of inverter-based renewable resources without their proper contribution to restoring frequency deviations following a disturbance. In this section, we show how applying the two-layer control architecture presented in Section \ref{sec:Optimal Control} yields a satisfactory overall grid response. We begin by optimizing the primary frequency control parameters as described in Section \ref{sec:Primary Control}. 
For the four operational regions, we select $\mu_1^{(i)} = 21$ and $\mu_2^{(i)} = \mu_3^{(i)} = 1$ for all $i \in \{C, E, W, S\}$, where \textit{C}, \textit{E}, \textit{W}, and \textit{S} denote the Central, Eastern, Western, and Southern regions, respectively.

For the constraints in (\ref{eq:Primary H}) and (\ref{eq:Primary D}), we adopt the values reported in \cite{Alhelou2025DistributedSystem}, namely $H_{min} = 0$, $H_{max} = 2.5$, $D_{min} = 0$, and $D_{max} = 1$. The droop gain in the KSA is restricted to $R_d = 0.04$ \cite{Alshehri2025MaintainingSupport}, and the minimum and maximum values of $K_d$ are set to $0$ and $1.2$, respectively. This results in the bounds of the constraint in (\ref{eq:Primary K}) being $\bar{K}_{min} = 0$ and $\bar{K}_{max} = 30$. 
After solving the optimization problem in (\ref{eq:Primary gamma}), the optimized parameters reported in Table \ref{tab:Parameters} are obtained.

\begin{figure}[h]
\centering
\includegraphics[width=\columnwidth]{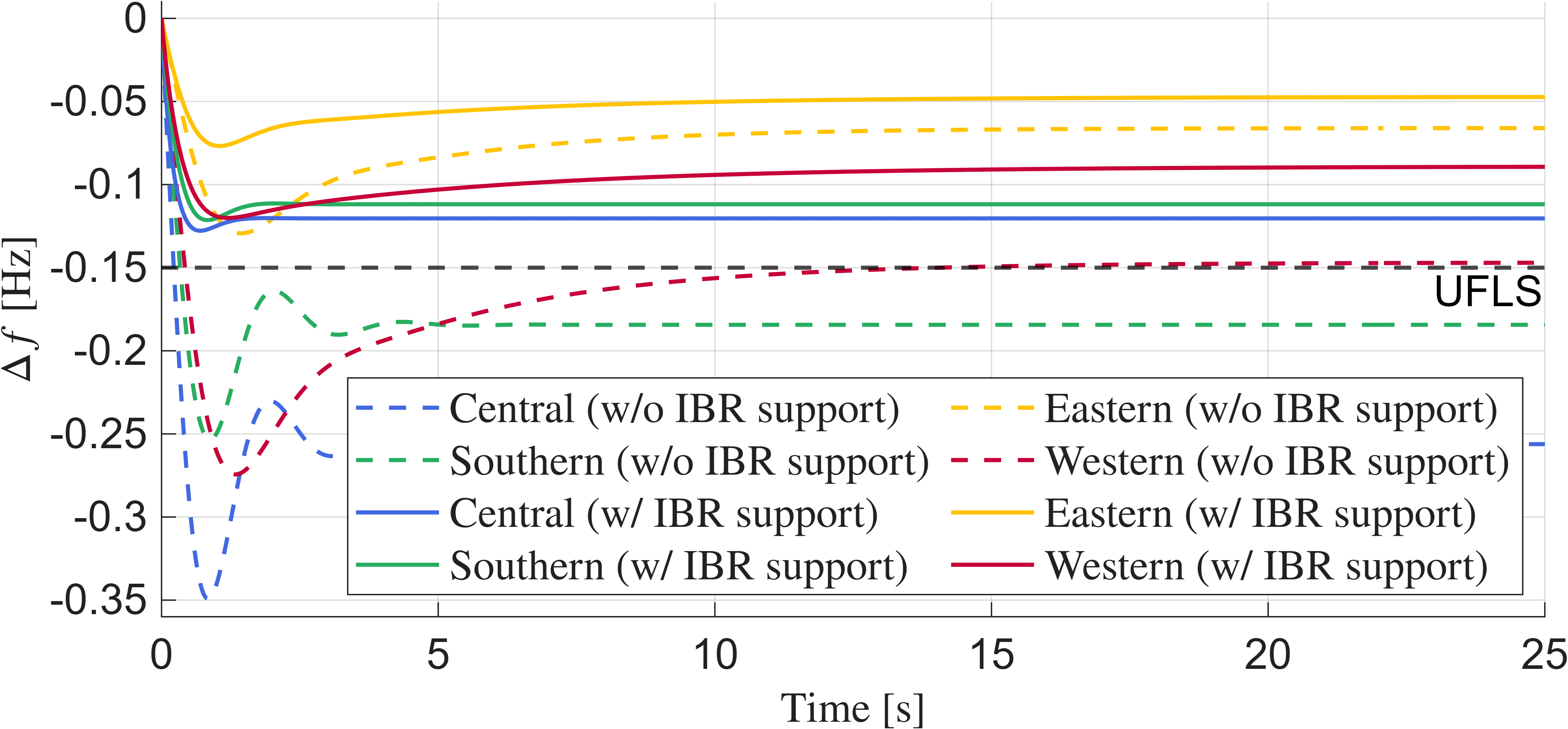}
\caption{Regional frequency deviation of the KSA grid following a 0.02 p.u. disturbance applied to each region, assuming no tie-line interconnections, shown with (solid) and without (dashed) IBR support under the projected 2030 RES penetration.}
\label{fig:OptimizedPrimaryResponse}
\end{figure}

Figure \ref{fig:OptimizedPrimaryResponse} shows the frequency response obtained using the optimized primary frequency control parameter derived from (\ref{eq:cost function primary})-(\ref{eq:Primary gamma}), assuming no tie-line interconnections among regions. The results demonstrate that the optimized parameters prevent the regional frequencies from reaching the UFLS threshold and reduce the steady-state frequency deviation.

\begin{table}[t!]
\renewcommand{\arraystretch}{1.6}
\centering
\caption{Optimized Primary Control Parameters}
\label{tab:Parameters}
\begin{tabular}{|c|c|c|c|c|}
\hline
\textbf{Optimal Parameter}                               & \textbf{Eastern} & \textbf{Central} & \textbf{Western} & \textbf{Southern} \\ \hline
$H_c^{(i)}$                        & $0.83$           & $1.19$           & $1.15$           & $1.17$            \\ \hline
$D_c^{(i)}$                        & $0.40$           & $0.44$           & $0.48$           & $0.47$            \\ \hline
$\bar{K}_d^{(i)}$ & $12.72$          & $15.44$          & $24.36$          & $18.97$           \\ \hline
$K_d^{(i)}$                        & $0.51$           & $0.62$           & $0.97$           & $0.76$            \\ \hline
$\gamma^{(i)}$       & $0.45$           & $0.68$           & $0.80$           & $0.80$            \\ \hline
\end{tabular}
\end{table}

\begin{figure}[ht]
    \centering

    \subfloat{\includegraphics[width=0.98\columnwidth]{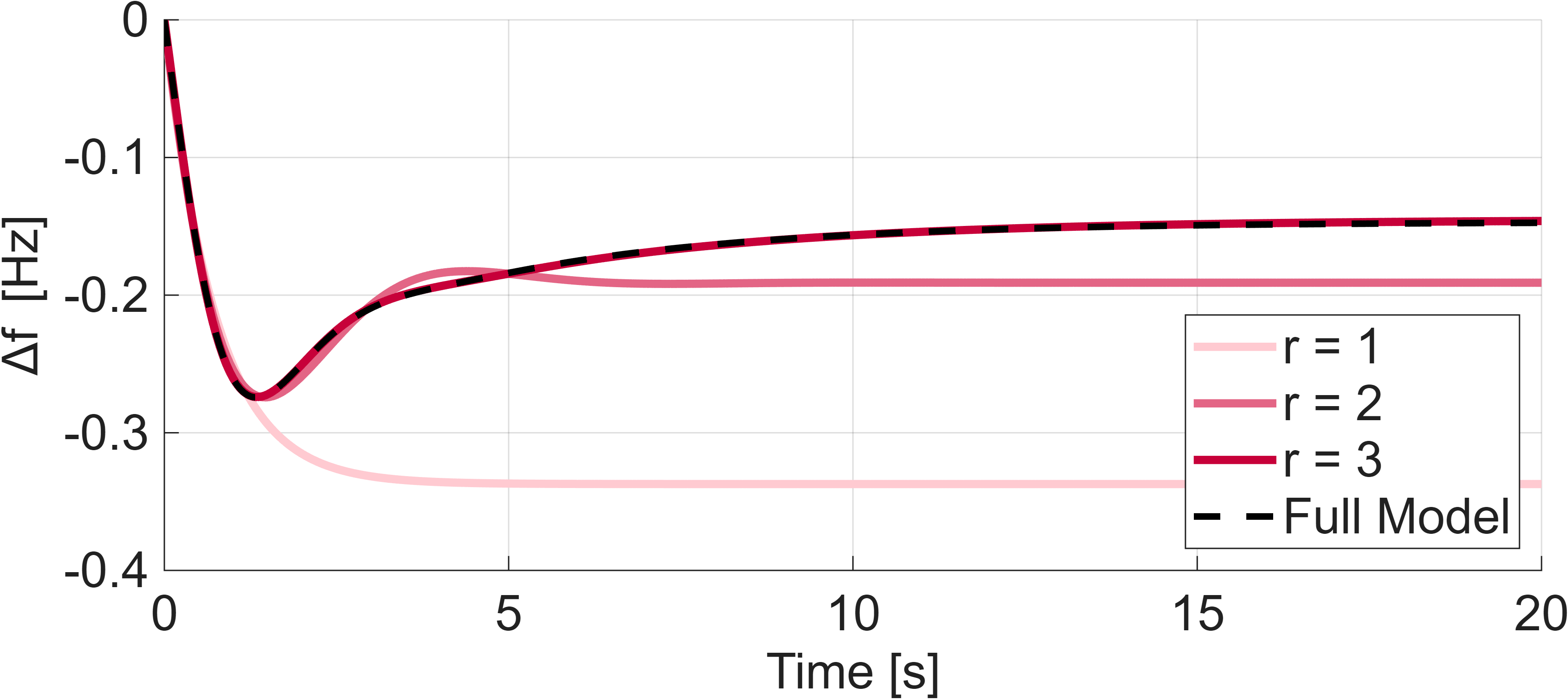}%
    }
    \vfil
    \subfloat{\includegraphics[width=0.47\columnwidth]{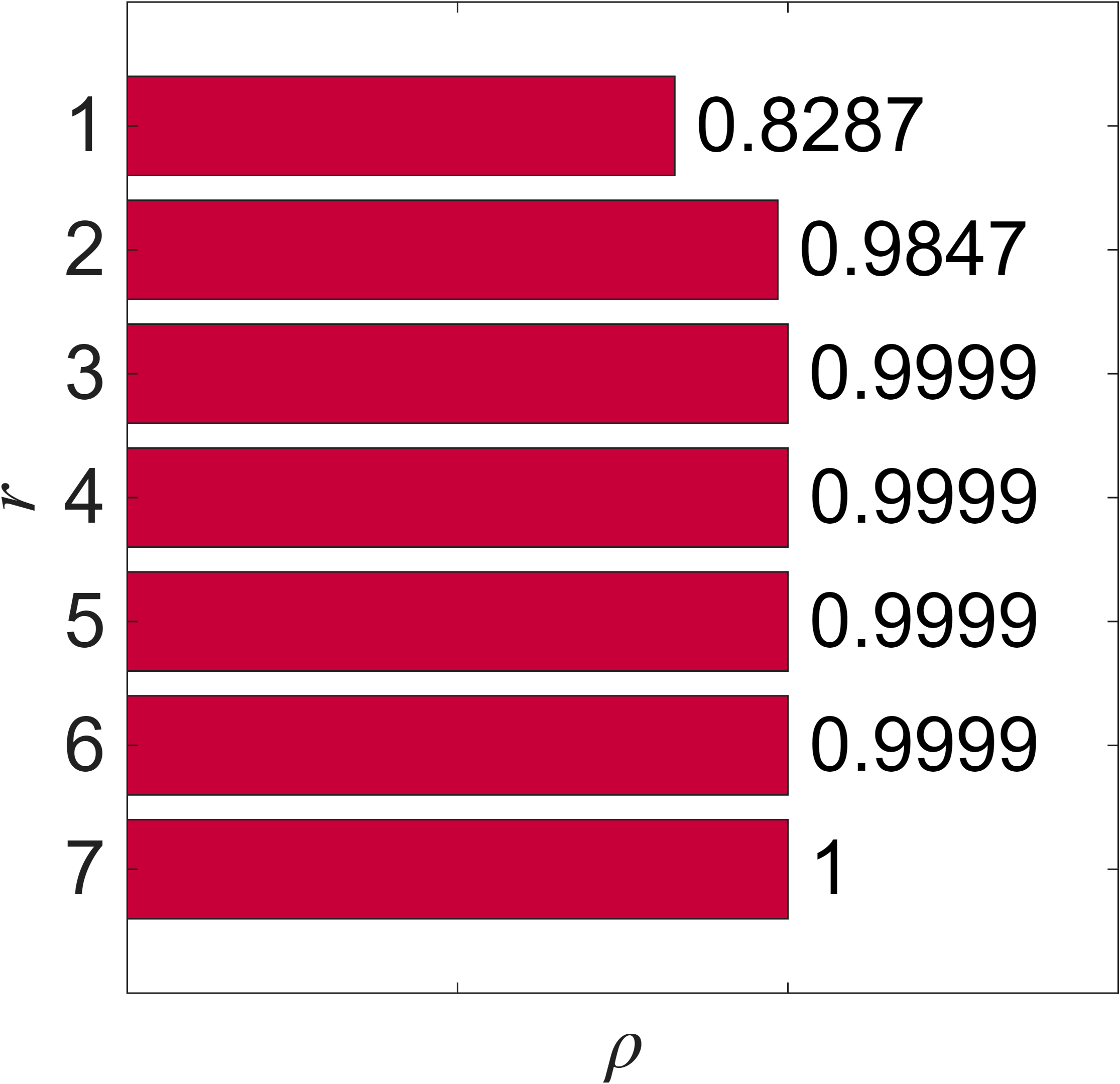}%
    }
    \hfil
    \subfloat{\includegraphics[width=0.47\columnwidth]{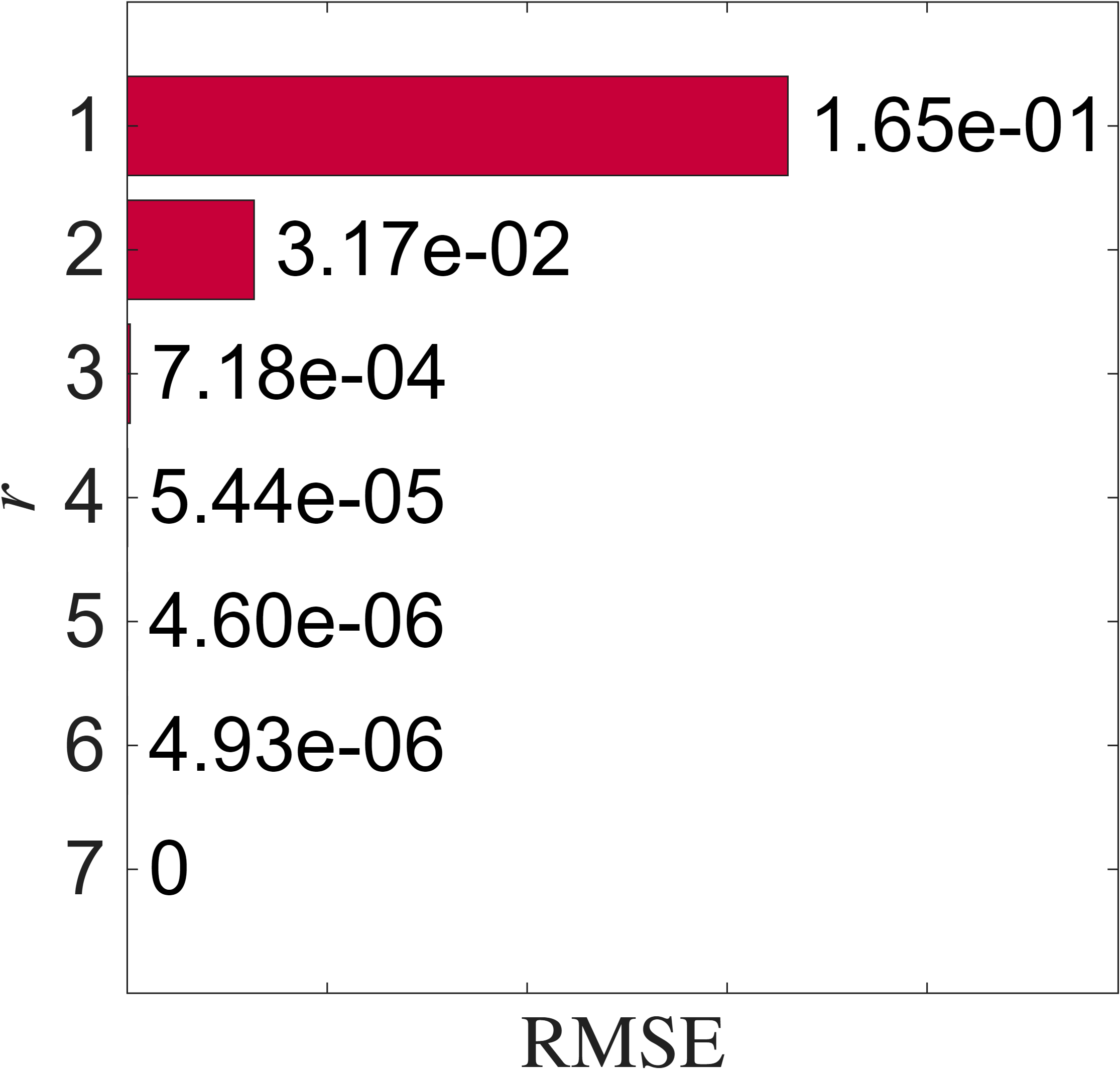}%
    }

    \caption{\textbf{Top:} Frequency response comparison of the full-order and one to three state reduced-order models for the Western region after a $0.02\ \text{p.u.}$ disturbance, assuming no tie-line interconnections. \textbf{Bottom-left:} Ratio $\rho$ as a function of reduced-order dimension $r$. \textbf{Bottom-right:} RMSE as a function of reduced-order dimension $r$.}
    \label{fig:HSV}
\end{figure}

We then proceed with the design of the MPC scheme. 
First, we derive the reduced-order observer as explained in Section \ref{sec:Observer}. The bottom-left panel in Fig. \ref{fig:HSV} shows the cumulative ratio $\rho(r)$, defined as
\begin{equation}\label{eq:rho ratio}
\rho(r) = \sum_{i=1}^{r} \sigma_i \left / \sum_{i=1}^{n_i} \sigma_i \right.
\end{equation}
for the Western operational region of the KSA power grid. This ratio quantifies the fraction of the total system energy captured by the first $r$ states. The top panel in Fig. \ref{fig:HSV} illustrates the corresponding time response for $r$ ranging from one to three, while the bottom-right panel shows the root mean square error (RMSE) for $r$ ranging from one to seven (the maximum for this area). As seen in the figure, selecting $r = 3$ captures approximately $99.99 \%$ of the system energy and yields a highly accurate prediction of the region's frequency response, as confirmed both qualitatively by the time response and quantitatively by the RMSE. A similar trend is observed for the other regions.

By selecting $r = 3$ for all regions, we systematically reduce the overall system order from $42$ to $28$ by computing a minimal realization of $G_i(s)$, and subsequently from $28$ to $16$ states using balanced truncation.

The remaining components of the observer and the MPC are designed as described in Sections \ref{sec:Observer} and \ref{sec:MPC}, using the following parameters: $Q_{\omega} = 10 I_{20}$, $R_{\nu} = I_{8}$, $Q = \operatorname{diag}(15, 15, 75, 15)$, $R = I_4$, $\eta_f = 1 \times 10^6 \mathbf{1}_4$, $\eta_u = 1 \times 10^3 \mathbf{1}_4$, $\underline{\Delta f} = -0.015 \ \text{Hz}$, $\overline{\Delta f} = 0.015 \ \text{Hz}$, $f_0 = 60 \ \text{Hz}$, $\overline{\text{RoCoF}} = 0.6 \ \text{Hz/s}$, $n_r = 3$, $\Delta t = 0.2 \ \text{s}$, $\underline{P}{TL} = -0.2 \ \text{p.u.}$, $\overline{P}{TL} = 0.2 \ \text{p.u.}$, $\overline{\Delta u} = 0.005 \ \text{p.u.}$, $\bar{P}_{IBR}^{\star} = 0.8$, $\underline{u} = -0.015\Lambda^{-1} \ \text{p.u.}$, and $\overline{u} = 0.015\Lambda^{-1} \ \text{p.u.}$


\subsection{Single-Area Disturbance Scenario}

\begin{figure}[h]
\centering
\includegraphics[width=0.96\columnwidth]{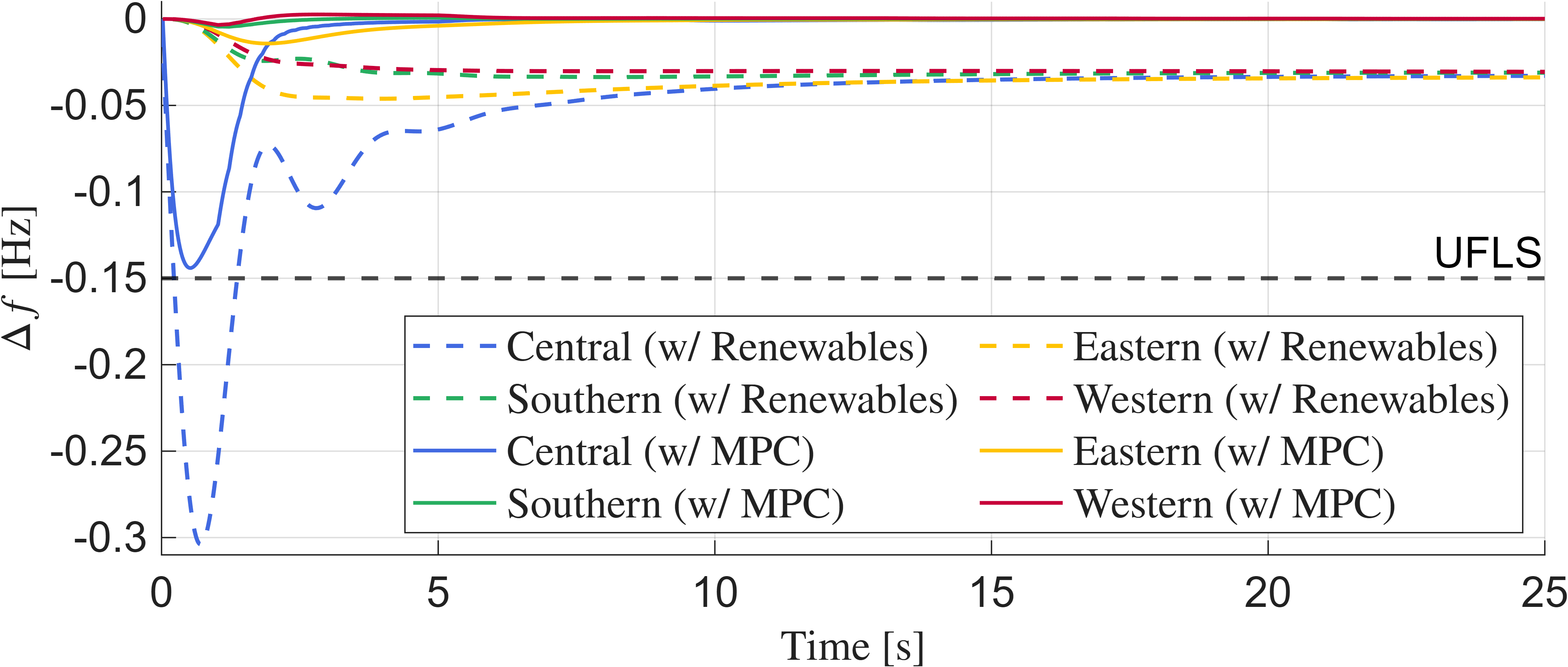}
\caption{Regional frequency deviations of the KSA grid following a $0.02\ \text{p.u.}$. disturbance in the Central region, shown with (solid) and without (dashed) IBR support using the proposed control architecture under the projected 2030 RES penetration.}
\label{fig:MPCResponse}
\end{figure}

\begin{figure}[h]
\centering
\includegraphics[width=0.96\columnwidth]{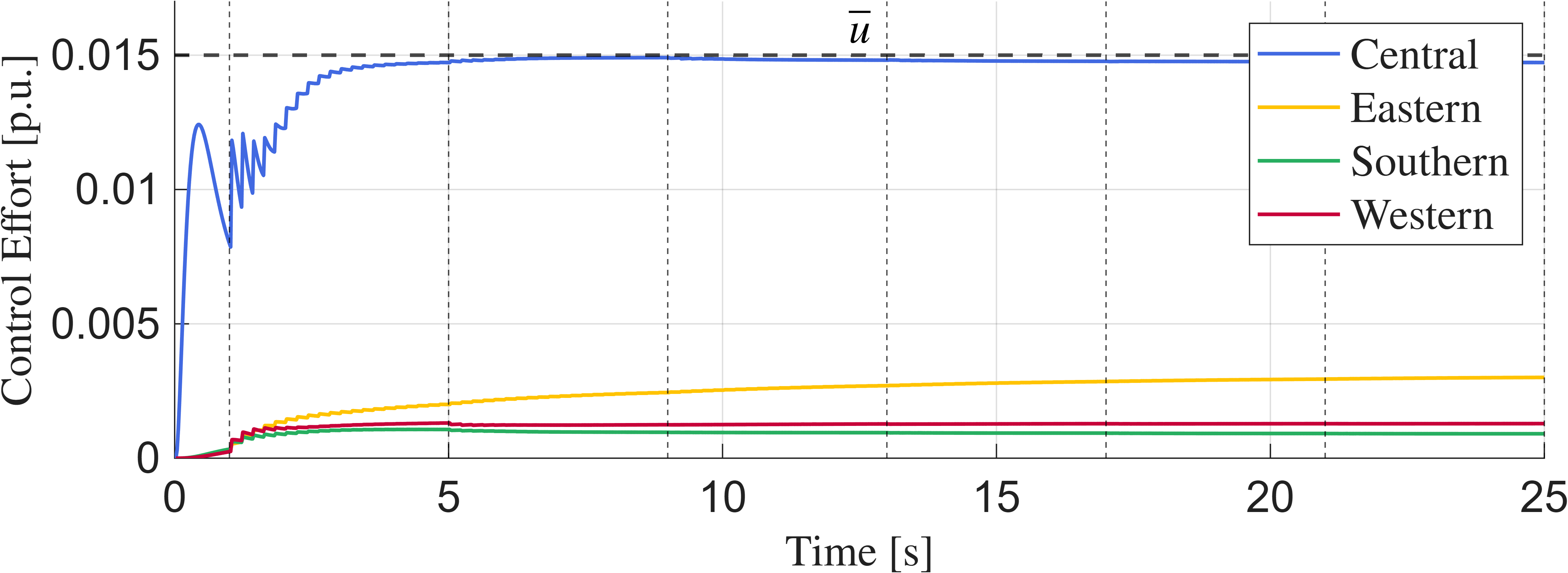}
\caption{Control effort in the four operational regions of the KSA grid following a $0.02\ \text{p.u.}$. disturbance in the Central region. Dashed vertical lines indicate the times at which the MPC issues control commands.}
\label{fig:MPCControl}
\end{figure}

The resulting frequency response and control effort following a disturbance of $0.02 \ \text{p.u.}$ in the Central region, with MPC and primary frequency control, are shown in Figs. \ref{fig:MPCResponse} and \ref{fig:MPCControl}, respectively. The initial estimated disturbance $\Delta \hat{P}(0)$ is $-0.018 \ \text{p.u.}$ in the Central region and $-0.001 \ \text{p.u.}$ in the Western region. The first MPC command is applied one second after the disturbance occurs in the Central region, as indicated by the vertical dashed lines in Fig. \ref{fig:MPCControl}. As shown, the proposed MPC strategy maintains the frequency above the UFLS limit and regulates it to the nominal value within approximately three seconds after the disturbance occurs. The control actions stay within scheduled power limits, despite the disturbance exceeding $\overline{u}$. 
This is achieved by optimally coordinating resources across different regions to compensate for the disturbance in the Central region.

\begin{figure}[ht]
\centering
\includegraphics[width=0.96\columnwidth]{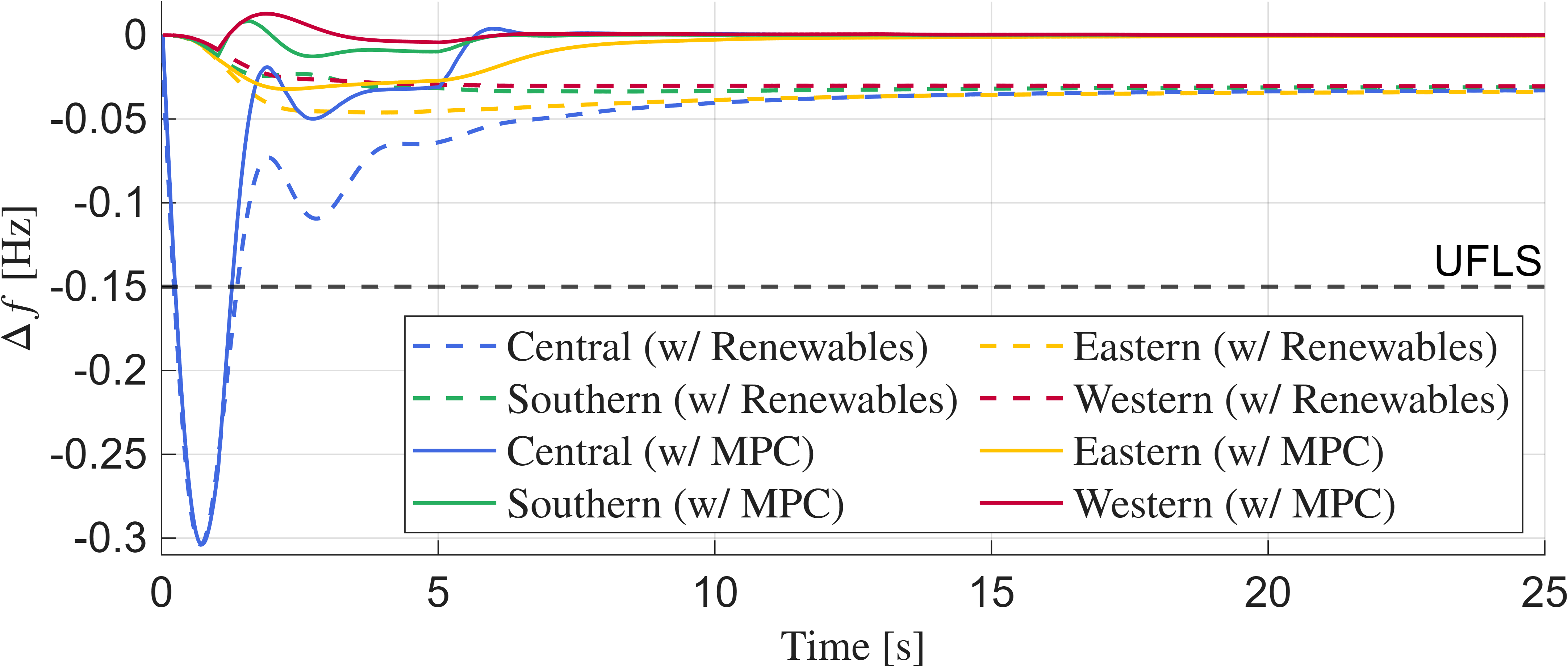}
\caption{Regional frequency deviations of the KSA grid following a $0.02\ \text{p.u.}$ disturbance in the Central region, shown with (solid) and without (dashed) IBR support using the described MPC but without primary frequency control, under the projected 2030 RES penetration.}
\label{fig:NoPrimary}
\end{figure}

To demonstrate the necessity of primary frequency control in the proposed design, we perform a simulation without it, as shown in Fig. \ref{fig:NoPrimary}. Although the MPC is still able to regulate the frequency to its nominal value, the first MPC command is computed only one second after the disturbance. As a result, a large frequency excursion occurs, crossing the UFLS limit. Furthermore, key performance metrics are degraded because of the absence of the primary control layer, as summarized in Table \ref{tab:Metrics}. In particular, the scheduled power constraint (\ref{eq:MPC power}) is violated due to the large initial frequency excursion. Since disturbances can occur at any time and cannot be synchronized with MPC updates, and given the control update rate constraints of the grid, the inclusion of primary frequency control is essential.

\begin{figure}[t]
\centering
\includegraphics[width=0.97\columnwidth]{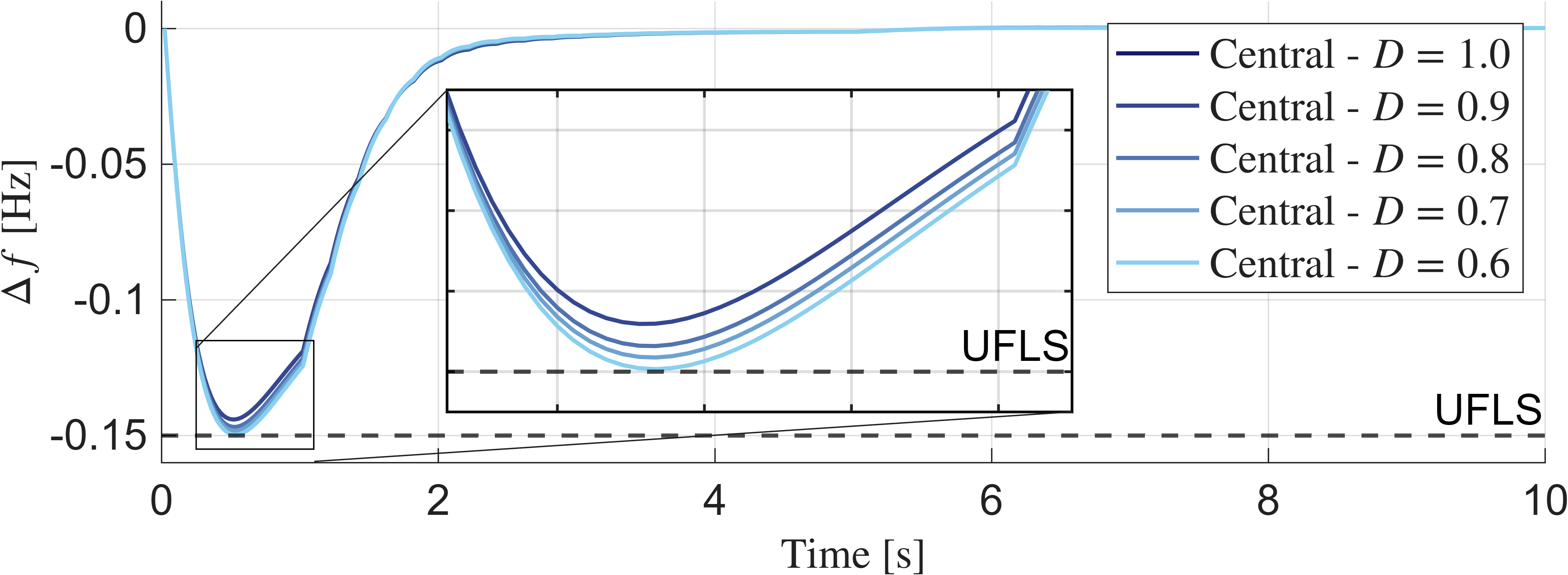}
\caption{Frequency deviation of the Central region following a $0.02 \ \text{p.u.}$ disturbance using the proposed control architecture for varying values of the damping coefficient $D_i$ between $0.6$ and $1.0$, under the projected 2030 RES penetration.}
\label{fig:Sensitivity}
\end{figure}

To assess the sensitivity of the proposed approach to the damping assumption discussed in Section \ref{sec:Sim No u}, we repeat the Central region disturbance scenario while varying the damping coefficient $D_i$ for all operational regions over the range $[0.6,1.0]$, representative of typical values reported in the literature. Fig. \ref{fig:Sensitivity} shows the resulting frequency deviation for the Central operational region for each value of $D_i$. We observed that even with a 40\% reduction in the damping coefficient, the frequency nadir increased by less than 4\% (with a minimum that does not go below -0.1497 Hz), indicating that the proposed control architecture is not sensitive to the specific choice of damping coefficient.

In the following simulations, we consider larger disturbances than the ones considered so far; to accommodate this, we increase $\underline{u}$ and $\overline{u}$ to $-0.02\Lambda^{-1} \ \text{p.u.}$ and $0.02\Lambda^{-1} \ \text{p.u.}$, respectively. All remaining parameters are kept as before. In addition, whenever a disturbance is detected, the observer's disturbance estimate is reinitialized with a randomly drawn value within $80-90 \%$ accuracy of the true magnitude. This reflects the fact that, in practice, the occurrence of a disturbance is readily detected from frequency measurements, even though its exact magnitude is not known a priori.

\subsection{Multi-Area Disturbance Scenario}

\begin{figure}[h]
\centering
\includegraphics[width=0.97\columnwidth]{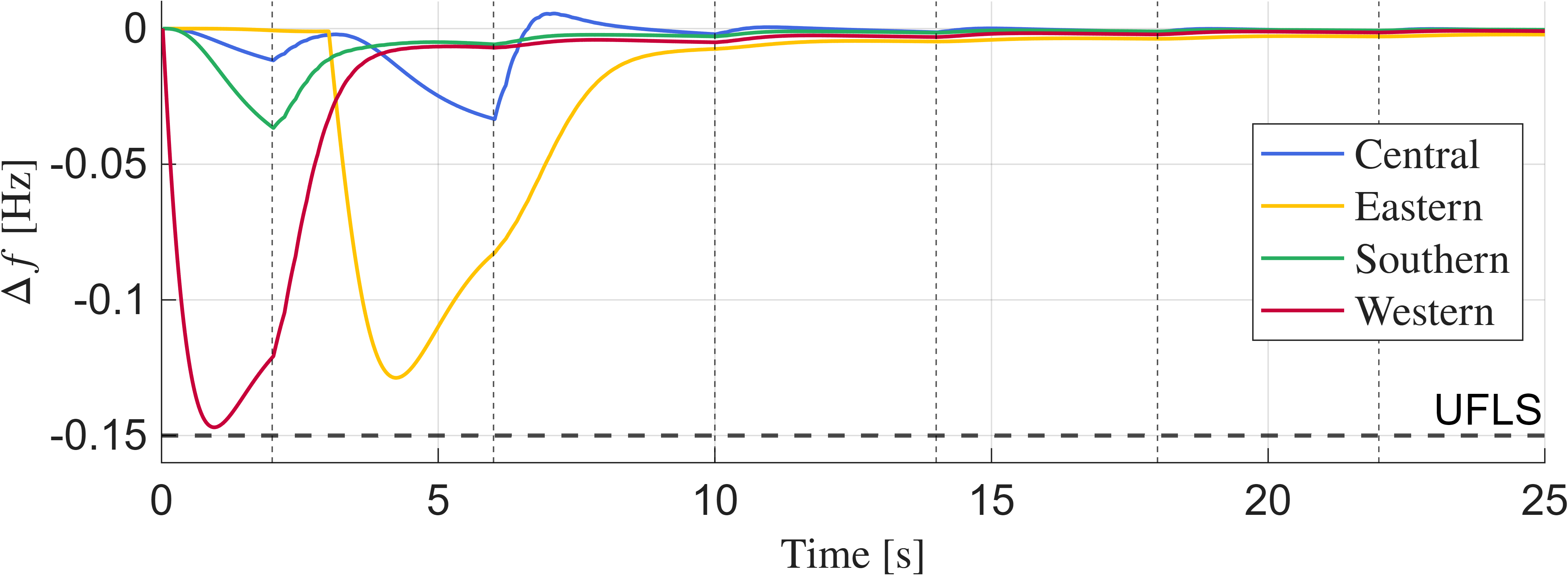}
\caption{Regional frequency deviations of the KSA grid following a $0.02 \ \text{p.u.}$ disturbance in the Western region at $t=0 \ \text{s}$ and a $0.025 \ \text{p.u.}$ disturbance in the Eastern region at $t=3 \ \text{s}$ using the proposed MPC under the projected 2030 RES penetration. Dashed vertical lines indicate the times at which the MPC issues control commands.}
\label{fig:Multiarea}
\end{figure}

We next evaluate the proposed architecture under a multi-area disturbance scenario, in which the Western and Eastern operational regions experience disturbances of $0.02 \ \text{p.u.}$ and $0.025 \ \text{p.u.}$, respectively, occurring at $t = 0$ and $t = 3$ seconds. The first MPC command is computed two seconds after the initial disturbance in the Western region, meaning that the second MPC update occurs three seconds after the disturbance in the Eastern region. Fig. \ref{fig:Multiarea} shows the resulting frequency response across all four operational regions, where the vertical dashed lines denote the times at which the MPC is computed. As shown in the figure, the frequency deviations remain within the UFLS limit, and the architecture successfully coordinates the different regions even though the frequency deviations from the first disturbance have not yet settled when the second disturbance occurs. The grid topology is evident from this response: since the Western region is interconnected with the Central and Southern regions, the effects of the initial disturbance are visible in these two regions. The Eastern region is interconnected only with the Central region, causing the Central region to also fluctuate when the disturbance in the Eastern region occurs. This scenario demonstrates that the proposed control architecture generalizes to disturbances of different magnitudes occurring in different regions and at different times relative to the SCADA update cycle. Note that we have chosen a conservative UFLS limit for this evaluation; for a less conservative value, the proposed approach can accommodate larger disturbances.

\subsection{Time-Varying Disturbance Scenario}

\begin{figure}[t]
\centering
\includegraphics[width=0.98\columnwidth]{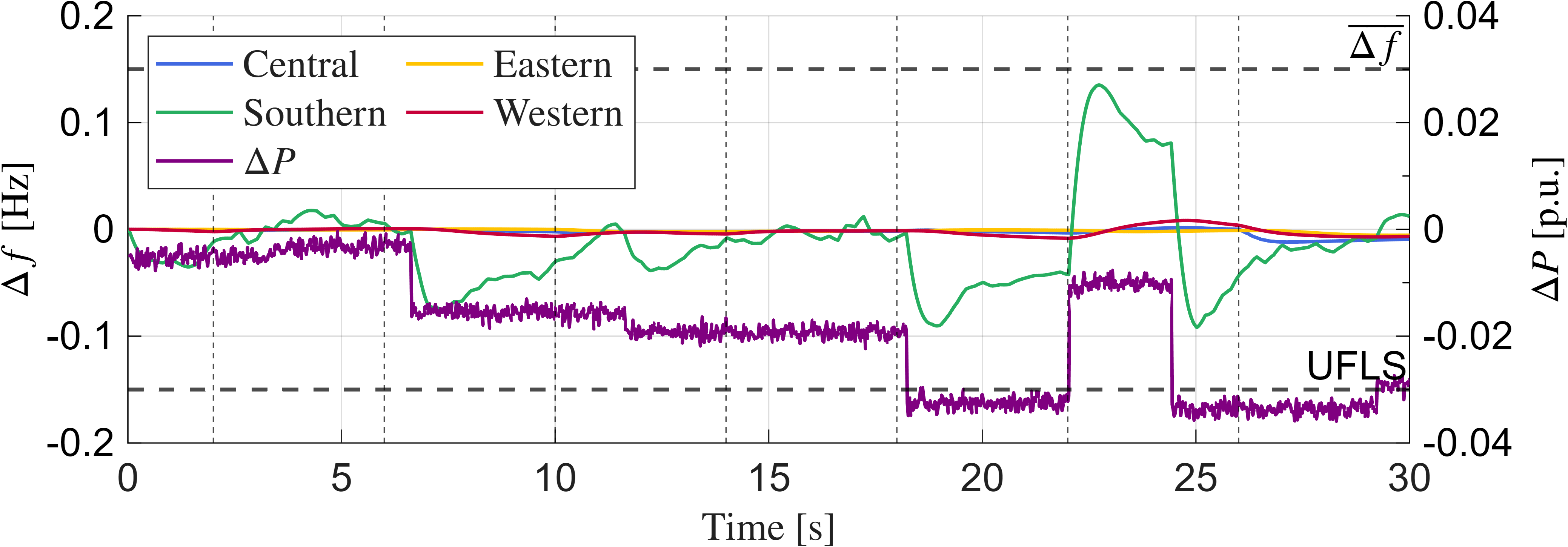}
\caption{Regional frequency deviations of the KSA grid and disturbance profile of a multilevel pseudorandom disturbance in the Southern region, under the projected 2030 RES penetration. Dashed vertical lines indicate the times at which the MPC issues control commands.}
\label{fig:PseudoRandom}
\end{figure}

Finally, to evaluate the proposed control architecture under a time-varying disturbance profile, we apply a multilevel pseudorandom signal with additive Gaussian noise $\mathcal{N}(0, 1\times10^{-7})$ to the Southern region, with step magnitudes drawn from a uniform distribution between $0.0$ and $0.04 \ \text{p.u.}$ and step durations drawn uniformly between $2$ and $8$ seconds, representative of the variability introduced by high renewable penetration \cite{HaesAlhelou2022DecentralizedShares}. Fig. \ref{fig:PseudoRandom} shows the resulting frequency response. As shown, the control architecture successfully keeps the frequency deviation within its limits and is robust to the added noise. A brief violation of the scheduled power constraint (\ref{eq:MPC power}) occurs at $t = 25$ s due to a particularly large disturbance realization; the slack-variable formulation allows the MPC to absorb this violation and swiftly restore constraint satisfaction at the next update. Furthermore, the mismatch between the SCADA update timing and the disturbance occurrence does not significantly affect the regulation capabilities of the proposed approach. This confirms that the proposed architecture remains effective under disturbance profiles that depart from the idealized step assumption motivating its design.

Table \ref{tab:Metrics} summarizes the frequency nadir, maximum RoCoF, settling time, control effort, and any constraint violations across all simulated scenarios.
\begin{table*}[t]
\centering
\caption{Performance Metrics Across Simulation Scenarios}
\label{tab:Metrics}
\begin{tabular}{lccccc}
\hline
\multicolumn{1}{|c|}{Scenario}                     & \multicolumn{1}{c|}{$\Delta f_\text{nadir}$} & \multicolumn{1}{c|}{Maximum RoCoF} & \multicolumn{1}{c|}{Settling Time$^2$}   & \multicolumn{1}{c|}{Control Effort} & \multicolumn{1}{c|}{Constraint Violations}  \\ \hline
\multicolumn{1}{|l|}{Central, baseline (Fig. \ref{fig:MPCResponse})}            & \multicolumn{1}{c|}{-0.1441}                                                & \multicolumn{1}{c|}{0.7184}        & \multicolumn{1}{c|}{3.3333}          & \multicolumn{1}{c|}{0.5437}         & \multicolumn{1}{c|}{No}                     \\ \hline
\multicolumn{1}{|l|}{Central, no primary (Fig. \ref{fig:NoPrimary})}          & \multicolumn{1}{c|}{-0.3039}                                                & \multicolumn{1}{c|}{0.7223}        & \multicolumn{1}{c|}{5.5111}          & \multicolumn{1}{c|}{0.5764}         & \multicolumn{1}{c|}{Yes, Constraints (\ref{eq:MPC power})} \\ \hline
\multicolumn{1}{|l|}{Multi-area (Western/Eastern) (Fig. \ref{fig:Multiarea})} & \multicolumn{1}{c|}{-0.1470/-0.1287}                                        & \multicolumn{1}{c|}{0.3875/0.2025} & \multicolumn{1}{c|}{11.0444/19.9111} & \multicolumn{1}{c|}{1.2821}         & \multicolumn{1}{c|}{No}                     \\ \hline
\multicolumn{1}{|l|}{Southern, pseudorandom (Fig. \ref{fig:PseudoRandom})}       & \multicolumn{1}{c|}{-0.0934-0.1340$^1$}                                        & \multicolumn{1}{c|}{0.5742}        & \multicolumn{1}{c|}{NA}              & \multicolumn{1}{c|}{0.3072}         & \multicolumn{1}{c|}{Yes, Constraints (\ref{eq:MPC power})} \\ \hline
\multicolumn{6}{l}{$^1$Minimum-Maximum, $^2$Time required for the frequency deviation to enter and remain within $2\%$ of $\Delta f_{\text{nadir}}$.}     
\end{tabular}
\end{table*}


\section{Conclusions}\label{sec:Conclusions}

This paper proposed a hierarchical frequency control architecture for interconnected power grids with high penetration of inverter-based resources. The proposed architecture integrates primary and secondary frequency control while explicitly accounting for operational constraints, such as realistic control update rates and converter limits. Evaluated on a detailed reduced-order model representative of the KSA grid under the 2030 vision, the framework incorporates balanced truncation and a reduced-order Kalman–Bucy observer to enable accurate state and disturbance estimation using only measurable outputs. Simulation results confirm that the approach maintains frequency above the UFLS threshold, restores nominal frequency within acceptable time, and respects operational constraints, while highlighting the indispensable role of fast primary control in limiting transient deviations. Overall, the proposed architecture offers an implementable and scalable solution for frequency regulation in renewable-dominated grids. 

Future work will be directed toward robustness to stochastic disturbances that accurately model the variability in renewable generation, distributed MPC schemes, integration of grid-following and grid-forming converters, and nonlinear effects.

\section*{Acknowledgment}

We would like to acknowledge Faisal Almutairi for his contributions to the data in Tables \ref{tab:Current Mix KSA} and \ref{tab:Inertia KSA}.




\appendix

\subsection{Observer Detectability Proof}\label{appendix A}

We provide a sufficient condition for the detectability of the disturbance-augmented system in (\ref{eq:reduced order x})-(\ref{eq:reduced order y}) and show how this condition is satisfied in practice. The proof proceeds in three steps: for ease of exposition, we begin with two operational regions, and first show that the reduced-order system is detectable, and second, establish a sufficient condition for the detectability of the disturbance-augmented system. Third, we show how these results generalize to $N$ operational regions and argue that the sufficient condition is generically satisfied in practice.

We introduce the following lemma that will be useful throughout the proof.

\textbf{Hautus Lemma for Detectability \cite{Damm2007OnSystems}:} Let $A \in \mathbb{R}^{n \times n}$ and $C \in \mathbb{R}^{m \times n}$. Then the following are equivalent:
\begin{enumerate}
    \item The pair $(A,C)$ is detectable.
    \item There exists no $v \neq 0$ and $\lambda$ such that $Av = \lambda v$ and $Cv = 0$ with $\Re(\lambda) \geq 0$.
\end{enumerate}

Consider the reduced-order single-area state-space models for two operational regions
\begin{equation}
    \hat{\Xi}_1 = \left(\bar{A}_{11}^{(1)}, \bar{B}_{1}^{(1)},\bar{C}_{1}^{(1)}, 0\right)
\end{equation}
and
\begin{equation}
    \hat{\Xi}_2 = \left(\bar{A}_{11}^{(2)}, \bar{B}_{1}^{(2)},\bar{C}_{1}^{(2)}, 0\right)
\end{equation} 
obtained as described in Section \ref{sec:Observer}. The two operational regions are interconnected by an AC tie-line with synchronizing coefficient $T_{12}$. The total rated apparent power of operational regions $1$ and $2$ are given by $S^{(1)}_T$ and $S^{(2)}_T$, respectively.

It can be verified that the assembled reduced-order state-space model for this two-region grid is given by the following matrices
\begin{align}
    \hat{A} & = \begin{bmatrix}
        \bar{A}_{11}^{(1)} & 0 & -\bar{B}_{1}^{(1)} \\
        0 & \bar{A}_{11}^{(2)} & \bar{B}_{1}^{(2)} \\
        \alpha^{(1)}\bar{C}_{1}^{(1)} & -\alpha^{(2)}\bar{C}_{1}^{(2)} & 0 
    \end{bmatrix} \\
    \hat{B} & = \begin{bmatrix}
        \bar{B}_{1}^{(1)} & 0 \\
        0 &  \bar{B}_{1}^{(2)} \\
        0 & 0 
    \end{bmatrix} \\
    \hat{C} & = \begin{bmatrix}
        \bar{C}_{1}^{(1)} & 0 & 0\\
        0 & \bar{C}_{1}^{(2)} & 0\\
        0 & 0 & 1 
    \end{bmatrix}
\end{align}
where
\begin{equation}
    \alpha^{(i)} = 2\pi T_{12}\left/ S^{(i)}_T \right.
\end{equation}

First, we show that the pair $(\hat{A}, \hat{C})$ is detectable. We proceed by contradiction. Assume that the pair $(\hat{A}, \hat{C})$ is undetectable. Then by the Hautus Lemma there exists a vector $v = [v_1^T, v_2^T, v_T]^T \neq 0$ and $\lambda$ with $\Re(\lambda) \geq 0$ such that $\hat{A}v=\lambda v$ and $\hat{C}v = 0$. The output conditions imply that
\begin{align}
    \bar{C}_{1}^{(1)}v_1 & = 0 \label{eq:C cond 1}\\
    \bar{C}_{1}^{(2)}v_2 & = 0 \label{eq:C cond 2}\\
    v_3 & = 0 \label{eq:tieline condition}
\end{align}

Substituting (\ref{eq:tieline condition}) into the eigenvalue equation $\hat{A}v=\lambda v$ yields
\begin{equation}
    \left(\bar{A}_{11}^{(1)} - \lambda I \right)v_1 = 0, \quad
    \left(\bar{A}_{11}^{(2)} - \lambda I \right)v_2 = 0
\end{equation}

Combined with (\ref{eq:C cond 1}), (\ref{eq:C cond 2}), and $\Re(\lambda) \geq 0$, observability of each $\left(\bar{A}_{11}^{(i)}, \bar{C}_{1}^{(i)}\right)$ \cite{Antoulas2004ApproximationOverview} forces $v_1 = v_2 = 0$. Thus $v = 0$, a contradiction $\contradiction$. Therefore the pair $(\hat{A}, \hat{C})$ is detectable.

Next, we provide a sufficient condition for the detectability of the disturbance-augmented system. The disturbance-augmented system is
\begin{align}
    A_{\text{ext}} & = \begin{bmatrix}
        \hat{A} & \hat{B} \\ 
        0 & 0
    \end{bmatrix} \label{eq: dist_aug 1} \\
    C_{\text{ext}} & = \begin{bmatrix}
        \hat{C} & 0
    \end{bmatrix} \label{eq: dist_aug 2}
\end{align}

Since $\hat{A}$ is Hurwitz (the grid without IBRs is stable), the only eigenvalue of $A_{\text{ext}}$ with  $\Re(\lambda) \geq 0$ is $\lambda = 0$, introduced by the piecewise-constant disturbance model. Therefore, detectability of $(A_{\text{ext}}, C_{\text{ext}})$ reduces to verifying the Hautus condition at $\lambda = 0$ only.

We again proceed by contradiction. Assume the pair $(A_{\text{ext}}, C_{\text{ext}})$ is undetectable. Then by the Hautus Lemma there exists $v = \begin{bmatrix} v^T_x & v^T_{\Delta P} \end{bmatrix}^T \neq 0$ such that
\begin{align}
    A_{\text{ext}}v & = \hat{A}v_x + \hat{B}v_{\Delta P} = 0 \label{eq:aug 1} \\
    C_{\text{ext}}v & = \hat{C}v_x = 0 \label{eq:aug 2}
\end{align}

Since $\hat{A}$ is Hurwitz, it is invertible, so from (\ref{eq:aug 1}) we obtain $v_x=-\hat{A}^{-1}\hat{B}v_{\Delta P}$. Substituting into (\ref{eq:aug 2}) gives:
\begin{equation}
    -\left( \hat{C}\hat{A}^{-1}\hat{B} \right)v_{\Delta P} = 0
\end{equation}

Therefore, if $\hat{C}\hat{A}^{-1}\hat{B}$ has full rank (equal to the number of disturbances, in this case two), the only solution is $v_{\Delta P} = 0$, which in turn implies $v_x = 0$ and hence $v = 0$, a contradiction. We conclude that the disturbance-augmented system is detectable whenever $\hat{C}\hat{A}^{-1}\hat{B}$ has full rank. Note that this matrix is the static gain from disturbance to output.

For the general case of $N$ operational regions, the assembled reduced-order state-space model has the following matrices:
\begin{align}
    \hat{A} & = \begin{bmatrix}
        \operatorname{diag}\left(\bar{A}_{11}^{(1)}, \cdots, \bar{A}_{11}^{(N)}\right) & F_1 \\
        F_2 & 0 
    \end{bmatrix} \\
    \hat{B} & = \begin{bmatrix}
        \operatorname{diag}\left(\bar{B}_{1}^{(1)}, \cdots, \bar{B}_{1}^{(N)}\right) \\
        0 
    \end{bmatrix} \\
    \hat{C} & = \begin{bmatrix}
        \operatorname{diag}\left(\bar{C}_{1}^{(1)}, \cdots, \bar{C}_{1}^{(N)}\right) & 0\\
        0 & I 
    \end{bmatrix} \label{eq: measurement N}
\end{align}
where $F_1$ collects how each tie-line flow enters the $i$-th operational region's dynamics, $F_2$ maps local frequency states to tie-line flow derivatives, and the bottom-right identity matrix in (\ref{eq: measurement N}) reflects that all $L$ AC tie-line flows are measured.

The proof follows the same steps. For detectability of $(\hat{A}, \hat{C})$, assume by contradiction that there exists $v=\begin{bmatrix}
    v_1^T & \cdots & v_N^T & v_T^T
\end{bmatrix}^T \neq 0$ and $\lambda$ with $\Re(\lambda) \geq 0$ such that $\hat{A}v=\lambda v$ and $\hat{C}v = 0$. Since all tie-line flows are measured, the output conditions immediately imply $v_T = 0$. Substituting into the eigenvalue equation using $v_T = 0$ to eliminate the coupling term $F_1v_T$, the per-region block equations decouple to $\left(\bar{A}_{11}^{(i)} - \lambda I \right)v_i = 0$ and $\bar{C}_{1}^{(i)}v_i = 0$ for all $i \in \{1, \cdots, N\}$. Observability of each $\left(\bar{A}_{11}^{(i)}, \bar{C}_{1}^{(i)}\right)$ then forces $v_i = 0$ for every operational region, implying $v = 0$, a contradiction $\contradiction$. Hence $(\hat{A}, \hat{C})$ is detectable. The disturbance-augmented system is then as in (\ref{eq: dist_aug 1}) and (\ref{eq: dist_aug 2}), and the same argument as in the two-region case shows that it is detectable whenever $\hat{C}\hat{A}^{-1}\hat{B}$ has full rank.

Physically, each disturbance $\Delta P_i$ primarily affects the local frequency deviation $\Delta f_i$, with a diminishing effect on the frequency of neighboring operational regions attenuated by the tie-line coupling. As a result, the $i$-th column of $\hat{C}\hat{A}^{-1}\hat{B}$ has a dominant entry in the row corresponding to $\Delta f_i$, giving the matrix an approximately diagonally dominant structure. Linear independence of the columns (and hence the rank condition) is therefore generically satisfied for any grid in which the $N$ operational regions have non-identical parameters or non-identical interconnection strengths. The only pathological case would be $N$ areas with perfectly identical parameters and perfectly symmetric interconnections, so that every disturbance produces the same steady-state signature in the measurements. This does not occur in any realistic multi-area grid, where the operational regions differ substantially in size, generation mix, and inertia.


\subsection{State vector for the Kingdom of Saudi Arabia Power Grid}\label{app:A}

For the SFR model used in the simulation of the KSA grid, the state vector $x(t) \in \mathbb{R}^{42}$ comprises the following components:
\begin{itemize}
    \item Central region (states $x_1(t)$-$x_9(t)$):  
    State $x_1(t)$ represents the regional frequency deviation. States $x_2(t)$-$x_5(t)$ correspond to the gas turbine dynamics, where $x_2(t)$ is the mechanical power command generated by the governor, $x_3(t)$ is the fuel valve position, $x_4(t)$ is the gas volume injection, and $x_5(t)$ is the compressor discharge, which is proportional to turbine torque and power output \cite{Rowen1983SimplifiedTurbines}. States $x_6(t)$-$x_9(t)$ are analogous to $x_2(t)$-$x_5(t)$ but correspond to the CCGT model.

    \item Eastern region (states $x_{10}(t)$-$x_{21}(t)$):  
    State $x_{10}(t)$ denotes the regional frequency deviation. States $x_{11}(t)$-$x_{13}(t)$ describe the steam turbine dynamics, where $x_{11}(t)$ is the steam valve position, $x_{12}(t)$ is the steam pressure, and $x_{13}(t)$ is the reheat steam energy storage, which is proportional to power output \cite{Anderson1990AModel}. States $x_{14}(t)$-$x_{21}(t)$ represent gas turbine and CCGT dynamics, defined analogously to those in the central region.

    \item Southern region (states $x_{22}(t)$-$x_{26}(t)$):  
    State $x_{22}(t)$ corresponds to the regional frequency deviation, while states $x_{23}(t)$-$x_{26}(t)$ represent gas turbine dynamics, as described previously.

    \item Western region (states $x_{27}(t)$-$x_{38}(t)$):  
    States $x_{27}(t)$-$x_{38}(t)$ are analogous to those of the eastern region and include the regional frequency deviation as well as steam turbine, gas turbine, and CCGT dynamics.

    \item Tie-line power flows (states $x_{39}(t)$-$x_{42}(t)$):  
    These states represent tie-line power flows between the central and eastern regions, central and southern regions, central and western regions, and southern and western regions, respectively.
\end{itemize}

\end{document}